\documentclass[11pt,english]{article}

\usepackage[utf8]{inputenc} 

\usepackage{dsfont}

\usepackage[left=1in,right=1in,top=1in,bottom=1in]{geometry}
\usepackage{authblk}
\usepackage{enumitem}
\usepackage{ulem} 
\usepackage{empheq}

\usepackage{graphicx}
\usepackage{subcaption}
\usepackage{float}
\usepackage{wrapfig} 

\usepackage{xcolor}
\usepackage{color}
\definecolor{darkgreen}{rgb}{0,0.5,0}
\definecolor{darkblue}{rgb}{0,0,0.6}
\definecolor{purple}{rgb}{0.4,.2,0.7}

\usepackage{amsmath}
\usepackage{amssymb}
\DeclareMathOperator{\sgn}{sgn}
\usepackage{tensor} 
\usepackage{slashed}
\usepackage{mathtools}
\usepackage{leftidx}
\usepackage{esint}
\usepackage{amsfonts,amsthm,bm}
\usepackage{tikz}
\usetikzlibrary{decorations.markings}
\tikzset
  {decoration=
     {markings,
      mark=at position 0.5 with {\arrow{stealth}}
     },
   plain/.style={line width=0.8pt},
   arrow/.style={plain,postaction=decorate}
  }
  
\usepackage{physics}
\usepackage{feynmf}
\usepackage{braket}

\usepackage{prettyref}
\usepackage[colorlinks=true,citecolor=darkgreen,linkcolor=purple,urlcolor=purple,breaklinks]{hyperref}
\usepackage{xurl} 

\usepackage[numbers,sort&compress]{natbib}

\newcommand{\nn}{\nonumber}

\def\Xint#1{\mathchoice
   {\XXint\displaystyle\textstyle{#1}}%
   {\XXint\textstyle\scriptstyle{#1}}%
   {\XXint\scriptstyle\scriptscriptstyle{#1}}%
   {\XXint\scriptscriptstyle\scriptscriptstyle{#1}}%
   \!\int}
\def\XXint#1#2#3{{\setbox0=\hbox{$#1{#2#3}{\int}$}
     \vcenter{\hbox{$#2#3$}}\kern-.5\wd0}}

\def\dashint{\Xint-}

\numberwithin{equation}{section}
\numberwithin{figure}{section}
\numberwithin{table}{section}

\begin{document}

\title{\LARGE\textsc{Real-time observables in de Sitter thermodynamics}}  


\author[a]{\vskip1cm \normalsize Manvir Grewal}
\affil[a]{ \it  Center for Theoretical Physics, Columbia University, New York, NY 10027, USA}
\author[b,c,d]{\normalsize Y.T.\ Albert Law}
\affil[b]{ \it \normalsize	 Center for the Fundamental Laws of Nature, Harvard University, Cambridge, MA 02138, USA}
\affil[c]{ \it \normalsize	 Black Hole Initiative, Harvard University, Cambridge, MA 02138, USA}
\affil[d]{ \it \normalsize	 Leinweber Institute for Theoretical Physics at Stanford, 382 Via Pueblo, Stanford, CA 94305, USA}
\date{}
\maketitle

\begin{center}
	\vskip-10mm
	{\footnotesize \href{mailto: mg3978@columbia.edu}{ manvir.grewal@columbia.edu}\, ,\;  \href{mailto:ytalaw@stanford.edu}{ytalaw@stanford.edu}}  
\end{center}

\vskip10mm

\thispagestyle{empty}

\begin{abstract}

We study real-time finite-temperature correlators for free scalars of any mass in a $dS_{d+1}$ static patch in any dimension. We show that whenever the inverse temperature is a rational multiple of the inverse de Sitter temperature, certain Matsubara poles of the symmetric Wightman function disappear. At the de Sitter temperature, we explicitly show how the Lorentzian thermal correlators can all be obtained by analytic continuations from the round $S^{d+1}$. We establish the precise relation between the Harish-Chandra character for $SO(1,d+1)$ and the integrated spectral function, providing a novel dynamical perspective on the former and enabling generalizations. Furthermore, we study scalars with exceptional non-positive masses. We provide a physical picture for the distinctive structures of their characters. For the massless case, we perform a consistent static patch quantization, and find the unique $S^{d+1}$ correlator that analytically continues to the correlators in the quantum theory.

\end{abstract}

\newpage

\tableofcontents


\section{Introduction}

Recent years have witnessed significant progress in understanding the semiclassical description of gravity and matter within a static patch of de Sitter (dS) space. For example, extensive calculations of the quantum-corrected dS horizon entropy, using Euclidean gravitational path integrals \cite{Anninos:2020hfj,Law:2020cpj,David:2021wrw,Anninos:2021ihe,Anninos:2021ene,Anninos:2021eit,Grewal:2021bsu,Muhlmann:2022duj,Bobev:2022lcc,Castro:2023dxp,Castro:2023bvo}, significantly refine the original Gibbons-Hawking proposal \cite{Gibbons:1976ue}, paving the way for constraining candidate microscopic models.\footnote{In $dS_3$, the dS entropy up to the logarithmic correction is reproduced by a $T \bar T+\Lambda_2$-deformed CFT$_2$ \cite{Shyam:2021ciy,Coleman:2021nor}.} In parallel, considerable discussions have been devoted to properly understanding the dS horizon entropy in terms of the algebra of observables associated with a static patch \cite{Chandrasekaran:2022cip,Witten:2023xze}, stressing the role played by the observer \cite{Anninos:2011af}.

Anticipating connecting these approaches and generalizations to interacting theories and beyond, we explore an object that played a key role in \cite{Anninos:2020hfj}: the Harish-Chandra character $\chi(g)$, defined as a trace of a group element $g$ over the representation space associated with UIRs of the isometry group $SO(1,d+1)$ of $dS_{d+1}$. The character associated with the dS boost, $g=e^{-i \hat H t}$, admits an expansion in terms of quasinormal modes (QNMs) in the static patch:
\begin{align}\label{eq:character}
    \chi(t) 
    =\sum_z D_z \, e^{-iz|t|} \; .
\end{align}
Here $z$ labels QNM frequencies and $D_z$ their degeneracies. Interpreting the Fourier transform 
\begin{align}\label{introeq:doschar}
    \tilde\rho^\text{dS}(\omega)\equiv \int_{-\infty}^\infty \frac{dt}{2\pi}e^{i\omega t} \chi(t) 
\end{align}
as a spectral density, the thermal canonical partition function for free fields in a static patch can be rendered well-defined, and for the case of scalars and spinors agrees with the 1-loop determinant on $S^{d+1}$. In \cite{Law:2022zdq,Grewal:2022hlo}, these results were extended to BTZ and Nariai black holes, and the precise meaning of \eqref{introeq:doschar} was clarified  through a generalized Krein-Friedel-Lloyd formula \cite{ChaosBook},
\begin{align}\label{introeq:Krein}
    \tilde\rho^\text{dS}(\omega)=\frac{1}{2\pi i}\partial_\omega \sum_{l} D_l^d \left(\log \mathcal{S}_l(\omega) - \log \mathcal{S}^{\rm Rin} (\omega)\right) \; , 
\end{align}
where $\mathcal{S}_l(\omega)$ and $\mathcal{S}^{\rm Rin} (\omega)$ are the scattering phases respectively for the scattering problems associated with the wave equation on the static patch and a reference Rindler-like space. $l$ labels the spherical harmonics $Y_l$ on $S^{d-1}$ with degeneracy $D_l^d$. A short review of these ideas can be found in \cite{Law:2023ohq}.

While the group-theoretic object \eqref{eq:character} via the spectral density \eqref{introeq:doschar} offers insights into making sense of traces or partition functions, one might wonder if there is a physical picture of \eqref{eq:character} directly in terms of observables within the static patch that is not intrinsically tied with the full dS isometries $SO(1,d+1)$. For one thing, a static patch is only invariant under the subgroup $SO(1,1)\times SO(d)$; for another, the putative finite dimensionality of the microscopic Hilbert space \cite{Banks:2000fe} suggests that the description in terms of $SO(1,d+1)$ UIRs is likely an emergent one in the semi-classical limit \cite{Witten:2001kn,Parikh:2004wh}.

The thermal nature of the static patch \cite{PhysRevD.15.2738} together with the fact that QNMs correspond to poles for the retarded propagator \cite{Anninos:2011af} naturally suggests that the relevant objects are real-time {\it thermal} correlators in the static patch. Hence, we start with a general discussion of real-time thermal correlators for a generic massive scalar in section \ref{sec:staticLor}. We will see that the spectral or commutator function
\begin{align}\label{introeq:GC}
    G^C(t,x,\Omega|0,y,\Omega')\equiv \bra{0} \left[\hat \phi (t,x,\Omega),\hat \phi (0,y,\Omega')\right] \ket{0}\;, 
\end{align}
which is temperature independent for a free theory ($\ket{0}$ is the vacuum defined with respect to the static patch Hamiltonian), plays a distinguished role. To probe the dependence on the inverse temperature $\beta$, we also study the symmetric Wightman function, and find that a subset of the Matsubara poles disappears whenever $\beta$ is a rational multiple of the inverse dS temperature $\beta_{\rm dS} = 2\pi \ell_{\rm dS}$ ($\ell_{\rm dS}$ is the dS length). In section \ref{sec:thermalfromsphere}, we explicitly compute various Lorentzian correlators at $\beta=\beta_{\rm dS}$ and show how they can all be obtained from an appropriate analytic continuation of the $S^{d+1}$ correlator.

We begin in section \ref{sec:characterStaticPatchViewpoint} by reviewing the basic ideas for the formula \eqref{introeq:Krein}, pointing out the distinction between resonance poles (poles of the retarded function) and scattering poles (poles of the scattering phase). Then, with an analysis inspired by the spectral theory on infinite-area hyperbolic surfaces \cite{borthwick_spectral_2016}, we establish the strikingly simple relation between the Harish-Chandra character \eqref{eq:character} and the spectral function \eqref{introeq:GC} for any scalars:
\begin{align}\label{introeq:charGC}
    \boxed{\chi(t)  = 2i \frac{d}{dt} \tilde\tr \,  \hat{G}^C(t)} \;. 
\end{align}
Here $\hat{G}^C(t)$ is the spectral function understood as an integral operator with kernel \eqref{introeq:GC}. The trace $\tilde\tr$ is implemented by an integration over the spatial slice, regulated in a suitable way. We stress that the right hand side of \eqref{introeq:charGC} is defined without reference to the dS isometries, and in principle can be extended to interacting theories.

Finally, in section \ref{sec:exceptional}, we study exceptional scalars labeled by a non-positive integer 
\begin{align}\label{introeq:shiftdim}
    \Delta = -k\;, \quad k=0,1,\cdots \; .
\end{align}
Despite the fact that the $SO(1,d+1)$ UIRs corresponding to \eqref{introeq:shiftdim} generically appear in the multiparticle Hilbert spaces of massive fields \cite{Dobrev:1977qv,Repka}, their realizations in terms of local quantum field theories remain to be understood. For the $k=0$ (massless) case, we show that its static patch quantization works as in the massive case as long as we exclude the zero mode. We also find the unique $S^{d+1}$ correlator that analytically continues to the various correlators in this quantum theory. For the $k\geq 1$ (tachyonic) case, the causal correlators grow with time, indicating the need for a drastic modification in order to have a consistent quantization. Lastly, we give a physical picture for the peculiar structure of the Harish-Chandra characters associated with \eqref{introeq:shiftdim}, which share similar features with (partially) massless fields \cite{Anninos:2020hfj,Sun:2021thf}, which include the physically interesting examples of Maxwell or graviton.

\paragraph{Plan of the paper}

In section \ref{sec:staticLor}, we study Lorentzian thermal correlators in the static patch. In section \ref{sec:thermalfromsphere}, we work out the analytic continuation from $S^{d+1}$. In section \ref{sec:characterStaticPatchViewpoint} we establish the relation between the Harish-Chandra character and the spectral function. In section \ref{sec:exceptional} we study scalars with exceptional non-positive masses. We conclude with some remarks in section \ref{sec:discussion}. We review the canonical quantization for free massive scalars in static patch in appendix \ref{sec:canquan}, and some basic facts in thermal field theory in appendix \ref{sec:ThermalFT}. In appendix \ref{app:Rin2D}, we work out the spectral theory on 2D Rindler space.


\section{Real-time static patch correlators}\label{sec:staticLor}

We consider a free scalar of mass $m^2$,
\begin{align}\label{eq:scalaraction}
    S = -\frac{1}{2}\int \sqrt{-g}\left[ \left(\partial \phi\right)^2+m^2 \phi^2 \right] \;, 
\end{align}
on a static patch in $dS_{d+1}$ with $d\geq 1$, whose metric is most commonly expressed in the static form
\begin{align}
	ds^2 = -\left(\ell^2-r^2 \right) d t ^ { 2 } + \frac { d r ^ { 2 } } { 1 - \frac { r ^ { 2 } } { \ell^ { 2 } } } + r ^ { 2 } d \Omega^2 \; , \qquad 0\leq r<\ell \; . 
\end{align}
The observer is located at $r=0$, surrounded by a horizon at $r=\ell$, with temperature $T_{\rm dS}=\frac{1}{2\pi \ell}$. Throughout this work, we will extensively use the tortoise coordinate $r=\ell \tanh x$ instead, with which the metric becomes
\begin{align}\label{eq:tortoisemetric}
	ds^2 = \ell^2 \frac{-dt^2 +dx^2 +\sinh^2 x \, d\Omega^2}{\cosh^2 x}\; , \qquad 0\leq x<\infty \; .
\end{align}
The observer is mapped to $x=0$ while the horizon is pushed to $x\to \infty$. Near horizon, the metric takes the Rindler-like form
\begin{align}\label{eq:tortoisemetricnearhor}
	\frac{ds^2}{ \ell^2} \stackrel{x\to \infty}{\approx} \; 4 \, e^{-2x }\left( -dt^2 +dx^2\right) +\, d\Omega^2\; .
\end{align}
From now on, we set the dS length $\ell$ to 1. In this section, we focus on the case $m^2>0$, i.e. either principal- or complementary-series scalars \cite{Sun:2021thf}. We will study the case of exceptional scalars with $m^2$ of some non-positive integer values in section \ref{sec:exceptional}.

After we rescale $\phi$ and separate the angular dependence with spherical harmonics as\footnote{For $d\geq 3$, the spherical harmonics $Y_{l}(\Omega)$ on the unit round $S^{d-1}$ satisfy
\begin{align*}
    - \nabla_{S^{d-1}}^2 Y_l (\Omega)= l(l+d-2)Y_l (\Omega) \; , \qquad l\geq 0\;,
\end{align*}
with degeneracy $D_l^d=\frac{d+2 l-2}{d-2}\binom{d+l-3}{d-3}$. For $d=1$, the angular dependence is trivial, and the range of $x$ should be extended to $-\infty <x<\infty$. For $d=2$, the spherical harmonics are $Y_l(\varphi)=\frac{e^{il\varphi}}{\sqrt{2\pi}},l\in \mathbb{Z}$. It is straightforward to modify the following discussions to the latter two cases.}
\begin{align}\label{eq:separateang}
    \phi (t,x,\Omega) =  \xi_{l }(t,x)Y_{l}(\Omega) = \tanh^{\frac{1-d}{2}}x \, \bar \xi_l (t,x) Y_{l}(\Omega) \;, 
\end{align}
the Klein-Gordon equation $\left(-\nabla^2+m^2\right)\phi =0$ decomposes into an infinite set of 1D Schr\"{o}dinger-like equations labeled by the $SO(d)$ quantum number $l\geq 0$, 
\begin{align}\label{eq:scalareomsep}
    \left(\partial_t^2 -\partial_x^2 +V_l (x)\right) \bar\xi_{l }(t,x)=0 \; ,\qquad 0\leq x<\infty\; , 
\end{align}
with a simple P\"{o}schl-Teller-like potential
\begin{align}\label{eq:potential}
    V_l (x) = \frac{(d+2l-3)(d+2l-1)}{4\sinh^2 x} -\frac{(2\Delta-d+1)(2\Delta-d-1)}{4\cosh^2 x} \; .
\end{align}
Here $\Delta$ is related to the mass through $m^2 = \Delta \bar\Delta$, with $\bar \Delta \equiv d-\Delta$. Since we will often go back and forth between quantities that are related by a $\tanh^{\frac{1-d}{2}}x$ rescaling as in \eqref{eq:separateang}, we find it convenient to introduce the bar notation, which should not be confused with complex conjugation.  Near horizon ($x\to \infty$), \eqref{eq:scalareomsep} reduces to 
\begin{align}\label{eq:nearhordsKG}
    \left(\partial_t^2 -\partial_x^2 +4\left( \Delta+l-1\right)\left( \bar\Delta+l-1\right)e^{-2x}\right) \bar\xi_{l }(t,x)=0 \; ,
 \end{align}
 which takes the same form as the Klein-Gordon equation of a massive scalar on a 2D Rinder space, as reviewed in appendix \ref{app:Rin2D}.

With the ansatz for the normal modes
\begin{align}\label{eq:scalaransatz}
     \xi_l (t,x) = e^{-i \omega t} \, \psi_{\omega l }(x)= e^{-i \omega t} \tanh^{\frac{1-d}{2}}x\, \bar\psi_{\omega l }(x)  \;, \qquad \omega \in\mathbb{R} \; , 
\end{align}
a general solution to \eqref{eq:scalareomsep} is a linear combination 
\begin{align}\label{eq:scalarlincom}
    \psi_{\omega l }(x) = C_{\omega l }^{\rm n.} \psi^{\rm n.}_{\omega l }(x)+C_{\omega l }^{\rm n.n.}\psi^{\rm n.n.}_{\omega l }(x) \;.
\end{align}
Here the (non-)normalizable mode\footnote{For even $d$, these expressions for $\psi^{\rm n.}_{\omega l }$ and $\psi^{\rm n.n.}_{\omega l }$ are not independent, and $\psi^{\rm n.n.}_{\omega l }$ needs to be replaced by a more complicated function. However, $\psi^{\rm n.n.}_{\omega l }$ are not required in our subsequent analysis and thus this issue can be neglected.}
\begin{align} \label{eq:scalarSol}
    \psi^{\rm n.}_{\omega l }(x) = \psi^{\rm n.n.}_{\omega, 2-d-l }(x) = \frac{\sech^{i \omega }x \, \tanh ^{l}x}{\Gamma \left(\frac{d}{2}+l\right)} \,
   _2F_1\left(\frac{l+\Delta +i \omega }{2},\frac{l+\bar\Delta +i \omega }{2} 
   ;\frac{d}{2}+l;\tanh ^2x\right) 
\end{align}
is (ir)regular at the location of the observer ($x=0$):
\begin{align}\label{eq:nearorigin}
    \psi^{\rm n.}_{\omega l }(x\approx 0)\propto  x^l \; , \qquad \psi^{\rm n.n.}_{\omega l }(x\approx 0)\propto  x^{2-d-l} \;.
\end{align}
It is also useful to note that the solutions \eqref{eq:scalarSol} are invariant under $\omega\to-\omega$:
\begin{align}
    \psi^{\rm n.}_{\omega l }(x) = \psi^{\rm n.}_{-\omega l }(x)\; , \qquad \psi^{\rm n.n.}_{\omega, 2-d-l }(x) = \psi^{\rm n.n.}_{-\omega, 2-d-l }(x) \; .
\end{align}
Using the identity
\begin{align}\label{eq:2f1 form}
_2F_1 (a,b;c;z ) =&\frac{\Gamma(c)\Gamma(c-a-b)}{\Gamma(c-a)\Gamma(c-b)}\,_2F_1 \left(a,b;a+b-c+1;1-z \right)\nn\\
&+\frac{\Gamma(c)\Gamma(a+b-c)}{\Gamma(a)\Gamma(b)}(1-z)^{c-a-b}\,_2F_1 \left(c-a,c-b;c-a-b+1;1-z \right) 
\end{align}
(valid for $|\arg(1-z)| < \pi$), one finds that near the horizon ($x\to\infty$)
\begin{align}\label{near hor}
	\psi^{\rm n.}_{\omega l }(x)=\psi^{\rm n.n.}_{\omega ,2-d-l }(x)\stackrel{x\to \infty}{\approx} &B_{\omega l} \, e^{i\omega x } +B_{-\omega l} \, e^{-i\omega x } \; , 
 \qquad 
    B_{\omega l} = \frac{2^{-i\omega}\Gamma(i\omega)}{\Gamma\left( \frac{\Delta+l+i\omega}{2}\right) \Gamma\left( \frac{\bar\Delta+l+i\omega}{2}\right)} \; .
\end{align}

For a static patch observer not carrying any source, it is natural to set $C_{\omega l }^{\rm n.n.}=0$ and discard the solution $\psi^{\rm n.n.}_{\omega l }(x)$ that blows up at $x=0$.\footnote{In $d=1$, both solutions \eqref{eq:scalarSol} are normalizable, and should both be kept as physical excitations.} Viewing each normalizable mode $\psi^{\rm n.}_{\omega l }(x)$ as a harmonic oscillator (with the overall coefficient $C_{\omega l }^{\rm n.}$ fixed by normalization with respect to the Klein-Gordon inner product), one can construct a Fock space following the usual canonical quantization procedure: $\omega$ and $l$ label single-particle eigenstates; multiparticle states are labeled by the associated occupation numbers. We review this construction in appendix \ref{sec:canquan}.

\subsection{Causal Green functions}\label{sec:causalfn}

As a building block for other real-time finite-temperature correlators (reviewed in appendix \ref{sec:ThermalFT}), we study the retarded two-point function, which for the free scalar theory coincides with the retarded Green functions associated with the Klein-Gordon equation,
\begin{gather}
     \left(-\nabla_x^2+m^2\right) G^R(t;x,\Omega| y,\Omega')=\frac{1}{\sqrt{-g}}\delta(t) \delta(x-y)\delta^{d-1}(\Omega,\Omega') \; , \label{eq:Greeneom} \\
     G^R(t;x,\Omega| y,\Omega')= 0\; , \qquad t<0 \;,  \label{eq:GRcondition}
\end{gather}
where without loss of generality we have put one of the two points at $t=0$. The advanced Green function $G^A(t;x,\Omega| y,\Omega')$ is similarly defined but with the retarded condition \eqref{eq:GRcondition} replaced by the advanced condition that it is zero for $t>0$ instead. It is straightforward to show that the retarded function for \eqref{eq:scalareomsep}, i.e.
\begin{align}\label{eq:rescaledGeom}
      \left(\partial_t^2 -\partial_x^2 +V_l(x)\right)\bar G^R_l(t;x| y)=\delta(t) \delta(x-y)\; ,
\end{align}
is related to the full retarded function \eqref{eq:Greeneom} by
\begin{align}\label{eq:retardrelation}
    G^R(t;x,\Omega| y,\Omega')=\tanh^{\frac{1-d}{2}}x \,\tanh^{\frac{1-d}{2}}y \, \sum_{l=0}^\infty \bar G^R_l(t;x| y) \, Y_l (\Omega) \, Y_l (\Omega') \;.
\end{align}
Similarly, all other correlators $G$ have a rescaled version $\bar G_l$ for each fixed $l$. In writing sums like $\sum_{l=0}^\infty$, we implicitly sum over the degenerate spherical harmonics labeled by their magnetic quantum numbers as well, so for each $l$ in the sum there are $D_l^d$ terms. 

We will be interested in the spectral properties of the single-particle Hamiltonian, so it is natural to work in the Fourier space:
\begin{align}\label{eq:rescaledGeomFourier}
      \left(-\omega^2 -\partial_x^2 +V_l(x)\right) \bar{\mathcal{G}}^R_l(\omega;x| y)=\delta(x-y)\; .
\end{align}
Herein we use $G$ and $\mathcal{G}$ to denote correlators in the time ($t$) and frequency ($\omega$) domains respectively. Since the potential \eqref{eq:potential} exponentially decays as $x\to \infty$, $\bar{\mathcal{G}}^R_l(\omega;x| y)$ is meromorphic on the complex $\omega$-plane \cite{Ching:1995tj}. The retarded condition \eqref{eq:GRcondition} is implemented by requiring that in the inverse transform 
\begin{align}
    \bar G^R_l(t;x| y) = \int_C d\omega \, e^{-i\omega t}\bar{\mathcal{G}}^R_l(\omega ;x| y)\;,
\end{align}
the contour $C$ lies above all the poles of $\bar{\mathcal{G}}^R_l(\omega ;x| y)$. Because of time-reversal symmetry, the advanced function is simply related to the retarded function by $\bar{\mathcal{G}}_l^A(\omega;x|y)=\bar{\mathcal{G}}^R_l(-\omega;x|y)$. The advanced condition is implemented by requiring that in the inverse transform the contour lies below all the poles of $\bar{\mathcal{G}}^A_l(\omega ;x| y)$.

From the standard theory of Green functions for ordinary differential equations, we can immediately write down a formula for the retarded function \cite{skinner2014mathematical}
\begin{align}\label{eq:full scalar retarded}
    2\pi \bar{\mathcal{G}}^R_l(\omega;x|y) = 
    \frac{\bar\psi^{\rm n.}_{\omega l }(x)\bar\psi^{\rm Jost}_{\omega l }(y)}{W^{\rm Jost}_{l}(\omega)} \theta(y-x)+\frac{\bar\psi^{\rm n.}_{\omega l }(y)\bar\psi^{\rm Jost}_{\omega l }(x)}{W^{\rm Jost}_{l}(\omega)} \theta(x-y)\;.
\end{align}
Here $\bar\psi^{\rm n.}_{\omega l }(x)=\tanh^{\frac{ d-1}{2}}x\,\psi^{\rm n.}_{\omega l }(x)$ is the rescaled version of \eqref{eq:scalarSol}, and 
\begin{align}\label{eq:jostsol}
    \bar\psi^{\rm Jost}_{\omega l }(x) 
   =&2^{i\omega }\sech^{-i \omega }x \tanh ^{\frac{d-1}{2}+l}x \, _2F_1\left(\frac{l+\Delta -i \omega}{2},\frac{l+\bar\Delta-i \omega}{2};1-i \omega ;\sech ^2x\right)
\end{align}
is the so-called Jost solution that is purely outgoing
\begin{align}\label{eq:Jostasym}
    \bar\psi^{\rm Jost}_{\omega l }(x) \stackrel{x\to \infty}{\approx}  \, e^{i\omega x }
\end{align}
near the horizon. {For any 1D scattering problem of the form \eqref{eq:scalareomsep}, }the Jost function, i.e., the Wronskian between $\bar\psi^{\rm n.}_{\omega l }$ and $\bar\psi^{\rm Jost}_{\omega l }$,
\begin{align}
    W^{\rm Jost}_{l}(\omega) \equiv \bar\psi^{\rm Jost}_{\omega l }(x) \partial_x \bar\psi^{\rm n.}_{\omega l }(x)-\partial_x\bar\psi^{\rm Jost}_{\omega l }(x)  \bar\psi^{\rm n.}_{\omega l }(x) \;,
\end{align}
appearing in \eqref{eq:full scalar retarded}, is independent of $x$ and can be computed at the horizon: 
\begin{align}\label{eq:Wronskian}
    W^{\rm Jost}_{l}(\omega) =-2i \omega B_{-\omega l} = \frac{  2^{1+i\omega}\Gamma(1-i\omega)}{\Gamma\left( \frac{\Delta+l-i\omega}{2}\right) \Gamma\left( \frac{\bar\Delta+l-i\omega}{2}\right)}\; .
\end{align}
Note that the overall normalization \eqref{eq:scalarlincom} of $\bar\psi^{\rm n.}_{\omega l }(x)$ does not affect that of \eqref{eq:full scalar retarded}. The locations of the zeros of \eqref{eq:Wronskian}, which we refer to as the {\bf resonance} poles, 
\begin{align}\label{eq:resonancepoles}
    i\omega^\Delta_{nl} = \Delta + 2n+l \;, \qquad \text{or} \qquad i\omega^{\bar\Delta}_{nl} = \bar\Delta + 2n+l \; , \qquad 0\leq n\in \mathbb{Z} \;,
\end{align}
correspond to the poles of \eqref{eq:full scalar retarded} on the complex $\omega$-plane. Similarly, the poles of $\bar{\mathcal{G}}_l^A(\omega;x|y)$, i.e. the {\bf anti-resonance} poles, are given by the zeros of $W^{\rm Jost}_{l}(-\omega)$. 

The resonance poles \eqref{eq:resonancepoles}, as poles of the retarded function, are also known as the quasinormal modes (QNMs) on a static patch. An alternative definition of QNMs characterizes them as being regular at the observer's location and purely outgoing at the horizon. However, these two definitions are not equivalent—a point we will elaborate on in section \ref{sec:Smatrix}. To clarify this distinction and avoid confusion, we will consistently use the term `resonance poles.'

Before moving on, we also note that the {\it poles} of $W^{\rm Jost}_{l}(\omega)$ and $W^{\rm Jost}_{l}(-\omega)$, corresponding to the {\it zeros} of $\bar{\mathcal{G}}_l^R(\omega;x|y)$ and $\bar{\mathcal{G}}_l^A(\omega;x|y)$ respectively, are located at 
\begin{align}\label{eq:Matpoles}
    i\omega^{\rm dS}_k = k \in \mathbb{Z} \setminus \{ 0\} \; , 
\end{align}
the (anti-)Matsubara frequencies associated with the dS temperature $\beta_{\rm dS}=2\pi$.

\subsection{Analytic structures of worldline thermal correlators}

Starting from the causal functions, we can build all other real-time thermal correlators, as reviewed in appendix \ref{sec:ThermalFT}. Without loss of generality, we focus on the behaviors of the correlators near the observer's worldline, i.e. when $x,y\approx 0$. Expressing the answer in terms of a dS-invariant distance, the correlators at any two timelike separated points can be obtained.

\subsubsection{The spectral function}

\begin{figure}[H]
    \centering
   \begin{subfigure}{0.35\textwidth}
            \centering
            \includegraphics[height=4.0cm]{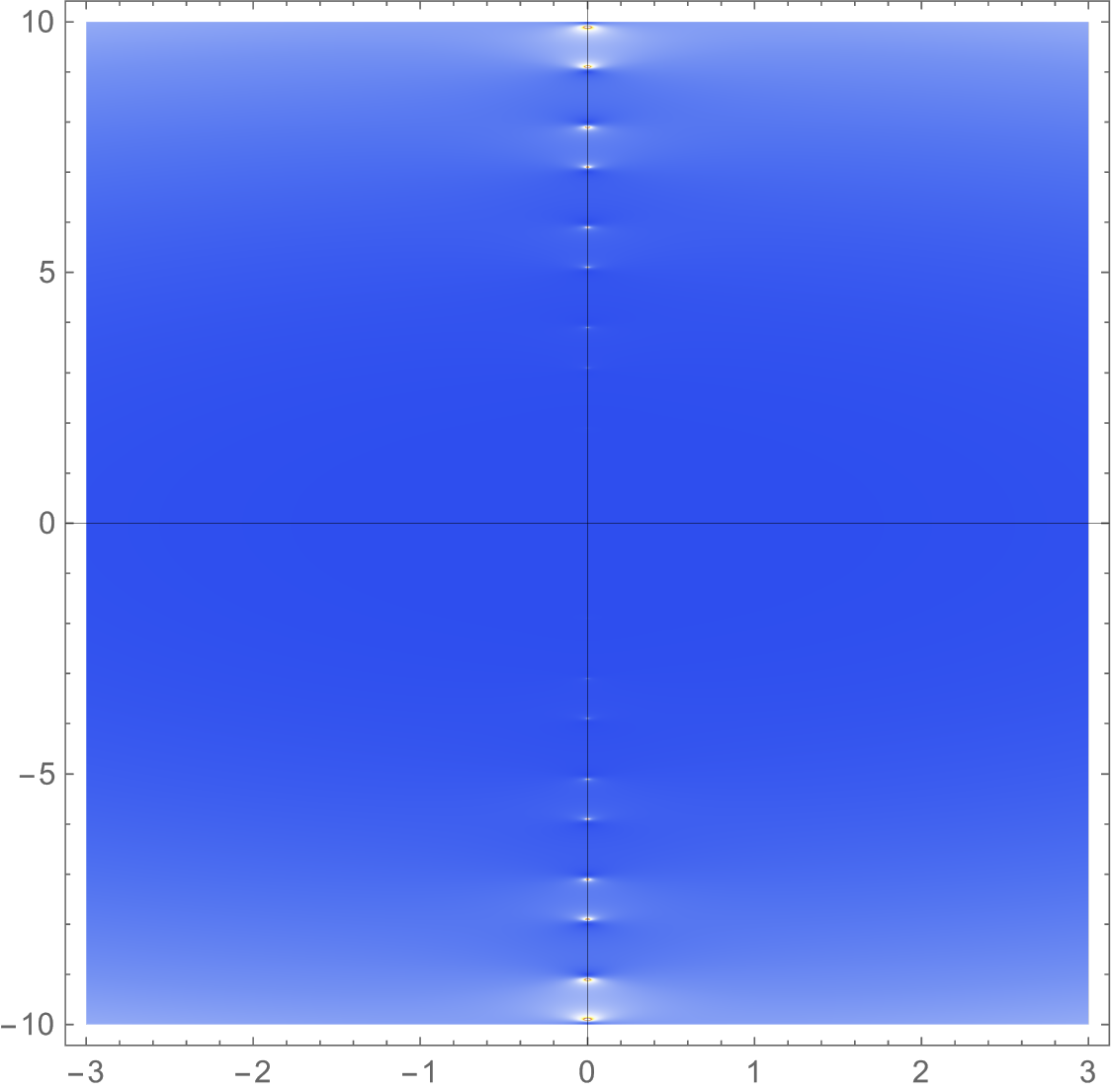}
            \caption[]%
            {{\small $\Delta=\frac{3}{2}+ 0.6$ (Complementary)}
             }    
        \end{subfigure}
        \hspace{1.2cm}
        \begin{subfigure}{0.35\textwidth}  
            \centering 
            \includegraphics[height=4.0cm]{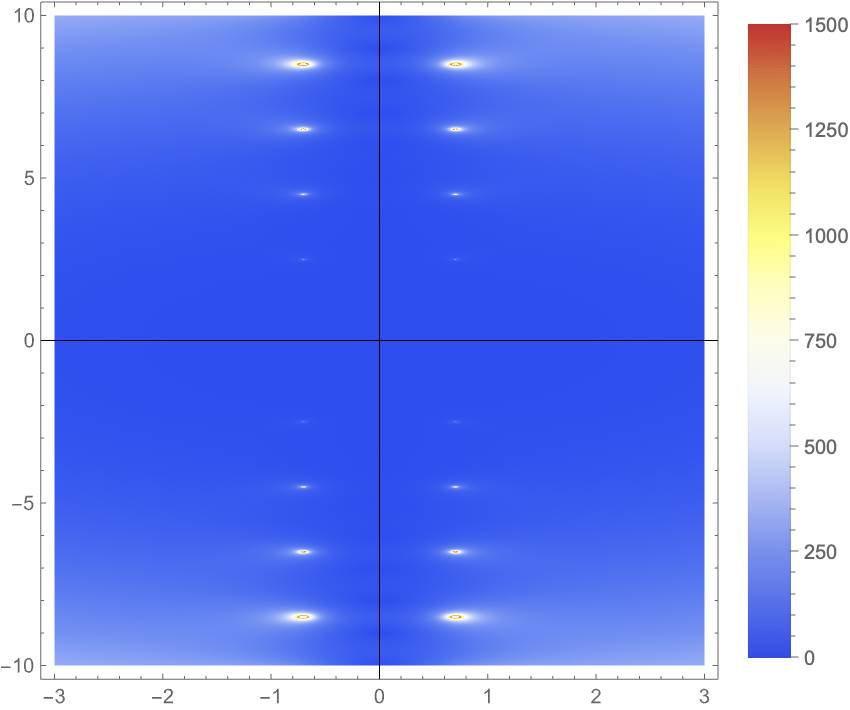}
            \caption[]%
            {{\small \centering  $\Delta=\frac{3}{2}+  0.7 i$ (Principal)}}    
        \end{subfigure}
        \caption{$\left|\mathcal{G}^C_{l=1}(\omega)\right|$ for a massive scalar in $dS_4$ in complex $\omega$-plane. { The white spots mark the poles.}}
        \label{pic:specfn}
\end{figure}

When $x,y\approx 0$, the full retarded function becomes 
\begin{align}\label{eq:worldtubeexpand}
    \mathcal{G}^R(\omega;x,\Omega|y,\Omega') \approx   \sum_{l=0}^\infty \mathcal{G}^R_l (\omega)  Y_l (\Omega)Y_l (\Omega') x^l y^l   \;,
\end{align}
where, according to a prescription motivated by the analogy with AdS/CFT \cite{Anninos:2011af}, $\mathcal{G}^R(\omega)$ is proportional to a ratio of the coefficients of the non-normalizable and normalizable modes such that the solution is purely outgoing at the horizon:
\begin{align}\label{eq:worldlineGR}
    \mathcal{G}^R_l (\omega) \propto \frac{B_{-\omega , 2-d-l}}{B_{-\omega l}} \; .
\end{align}
Alternatively, this can be obtained by taking $x,y\to 0$ in \eqref{eq:full scalar retarded}, which fixes the overall $\omega$-independent factor as well. The difference between the retarded and advanced functions gives the so-called spectral or commutator function: 
\begin{align}\label{eq:scalarspectral}
    \mathcal{G}^C_l (\omega) =-i\left( \mathcal{G}^R_l (\omega)-\mathcal{G}^A_l (\omega)\right)
    \propto \sinh \pi \omega \left| \Gamma\left( \frac{\Delta+l-i\omega}{2}\right) \Gamma\left( \frac{\bar\Delta+l-i\omega}{2}\right)  \right|^2 \; . 
\end{align}
Other than the (anti-)resonance poles (see Figure \ref{pic:specfn}) inherited from $\mathcal{G}^{R/A}_l (\omega)$, \eqref{eq:scalarspectral} has {\it zeros} at
\begin{align}\label{eq:spectralzeros}
    i\omega^{\rm dS}_{k} = k \; , \qquad k \in \mathbb{Z} \; ,
\end{align}
which are the (anti-)Matsubara frequencies associated with $\beta_{\rm dS}=2\pi$, including the $\omega=0$ mode. 
The spectral function \eqref{eq:scalarspectral} plays a particularly important role because it encodes all the information for the single-particle Hamiltonian, and is directly related to the Harish-Chandra character. We will elaborate on this in section \ref{sec:specchar}.

\subsubsection{Thermal symmetric Wightman function}

As reviewed in appendix \ref{sec:ThermalFT}, the causal functions (and thus the spectral function) for a free theory at finite inverse temperature $\beta$ are independent of $\beta$. To see temperature dependence, we must study {Wightman-type} correlators, such as the thermal symmetric function:
\begin{align}\label{eq:symWightman}
    G^{\rm Sym}_{\beta}(t)\equiv \frac{1}{Z(\beta)}\Tr \, \left[ e^{-\beta \hat H} \left( \frac{\hat \phi (t)\hat \phi (0)+\hat \phi (0)\hat \phi (t)}{2}\right) \right] \; , \qquad Z(\beta) \equiv \Tr \, e^{-\beta \hat H} \; ,
\end{align}
where we have suppressed spatial dependence. {Note that the ordinary thermal Wightman function, \begin{align}\label{eq:Wightman}
    G^{>}_{\beta}(t)\equiv \frac{1}{Z(\beta)}\Tr \, \left( e^{-\beta \hat H} \hat \phi (t)\hat \phi (0) \right) \; , 
\end{align}
can be reconstructed from the symmetric correlator together with the commutator function, so no information is lost. We will therefore focus on \eqref{eq:symWightman}.

In Fourier space, \eqref{eq:symWightman}} is related to the spectral function at a general inverse temperature $\beta$ {(see appendix \ref{sec:ThermalFT}):}
\begin{align}\label{eq:wightmanretarded}
    \mathcal{G}^{\rm Sym}_l (\beta,\omega) = \frac{1}{2}\coth \frac{\beta \omega}{2} \mathcal{G}^C_l (\omega) \propto \coth \frac{\beta \omega}{2} \sinh \pi \omega \left| \Gamma\left( \frac{\Delta+l-i\omega}{2}\right) \Gamma\left( \frac{\bar\Delta+l-i\omega}{2}\right)  \right|^2 \;.
\end{align}
The factor $\coth \frac{\beta \omega}{2}$ reflects the fact that the particles obey bosonic statistics. From \eqref{eq:wightmanretarded}, we see that the analytic structure of \eqref{eq:wightmanretarded} depends on whether $\beta$ is a rational multiple of $2\pi$. Analogous dependence has been pointed out at the level of free energies in \cite{Akhmedov:2021cwh}.

\paragraph{Generic $\beta$}

\begin{figure}[H]
    \centering
   \begin{subfigure}{0.3\textwidth}
            \centering
            \includegraphics[height=4.0cm]{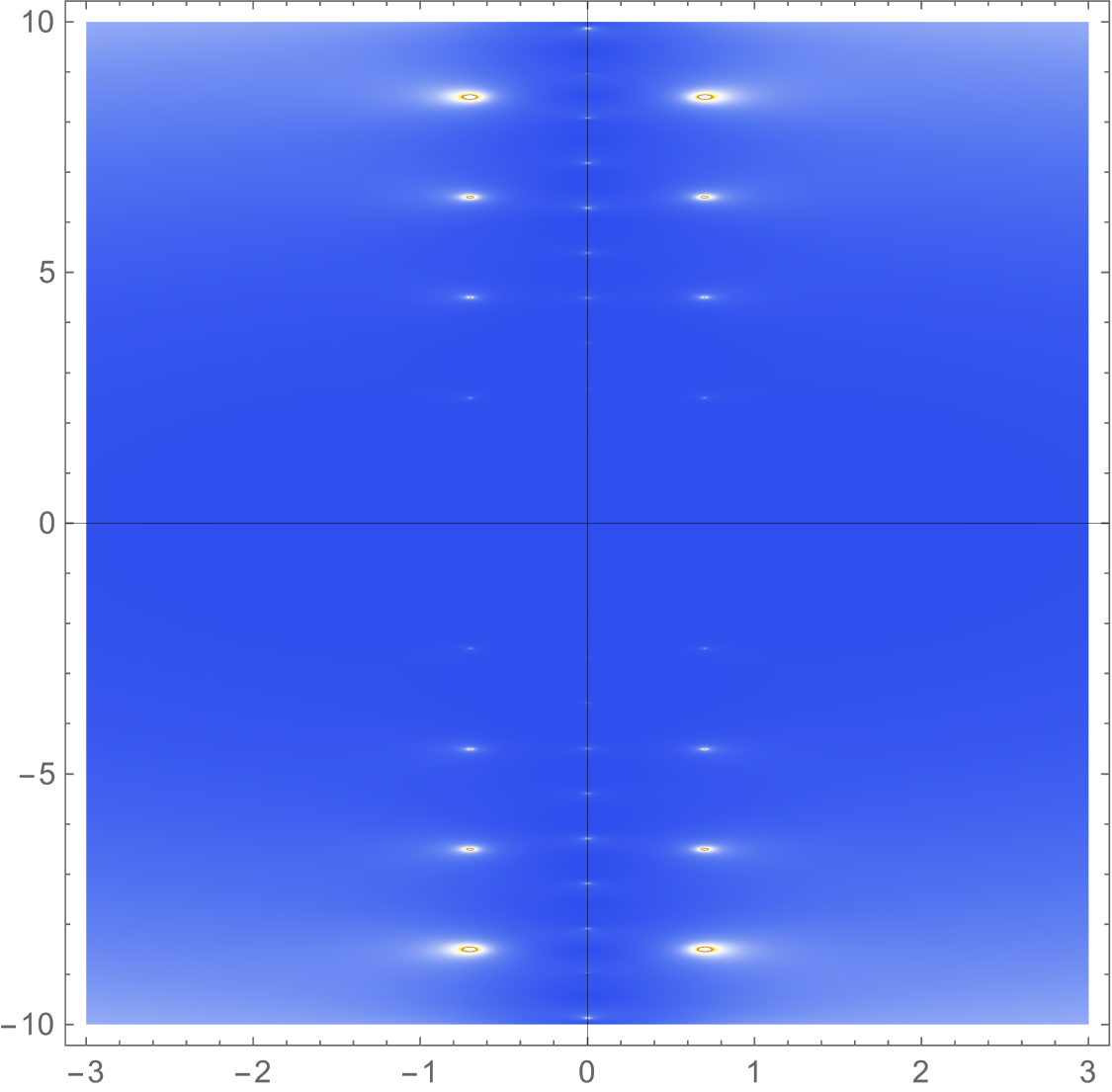}
            \caption[]%
            {{\small $\beta=5$}
             }    
             \label{fig:symgeneric}
        \end{subfigure}
        \hspace{0.3cm}
        \begin{subfigure}{0.3\textwidth}  
            \centering 
            \includegraphics[height=4.0cm]{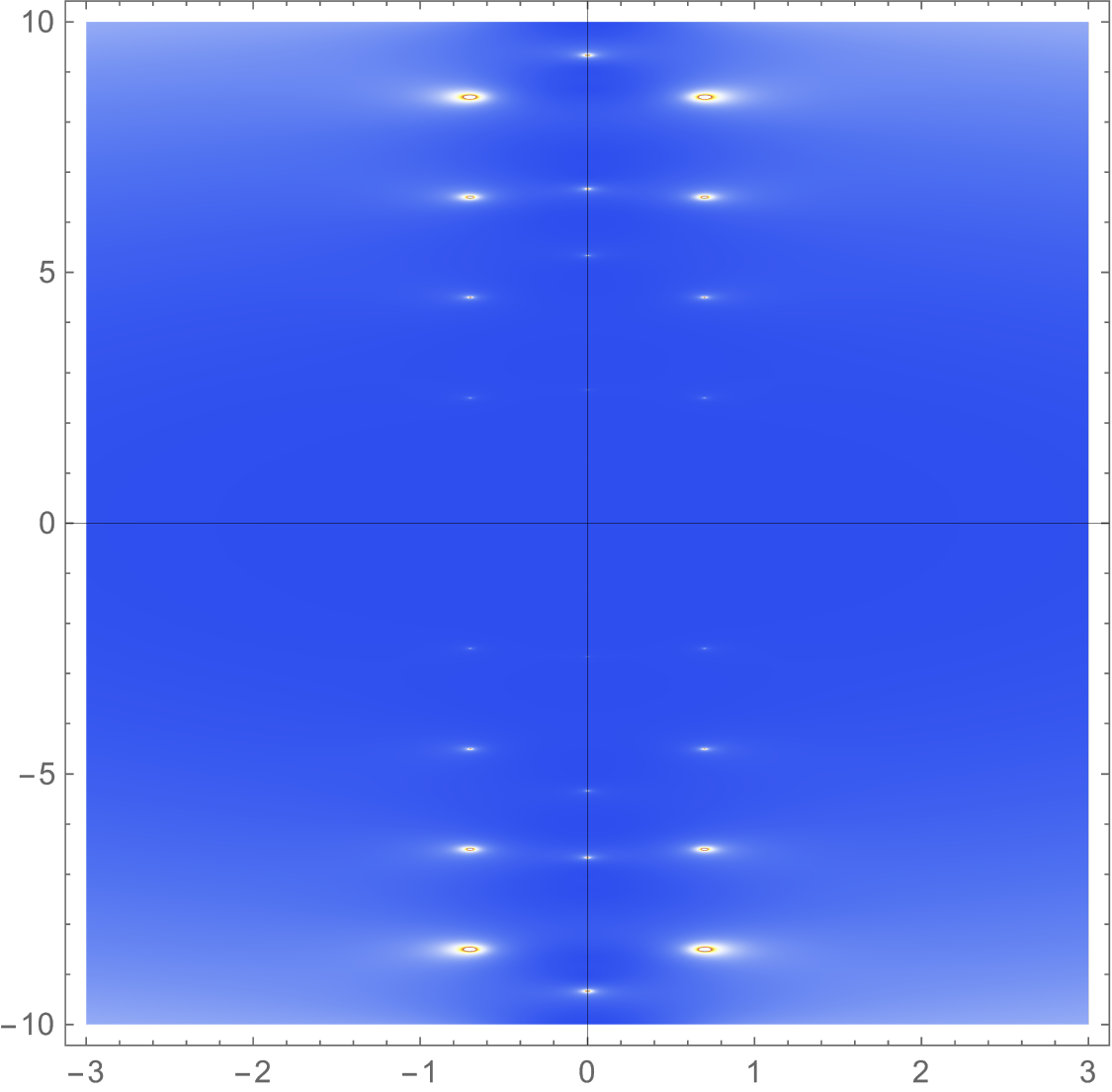}
            \caption[]%
            {{\small \centering  $\beta=\frac{3\pi}{2}$}}
            \label{fig:symspecial}
        \end{subfigure}
        \hspace{0.3cm}
        \begin{subfigure}{0.3\textwidth}  
            \centering 
            \includegraphics[height=4.0cm]{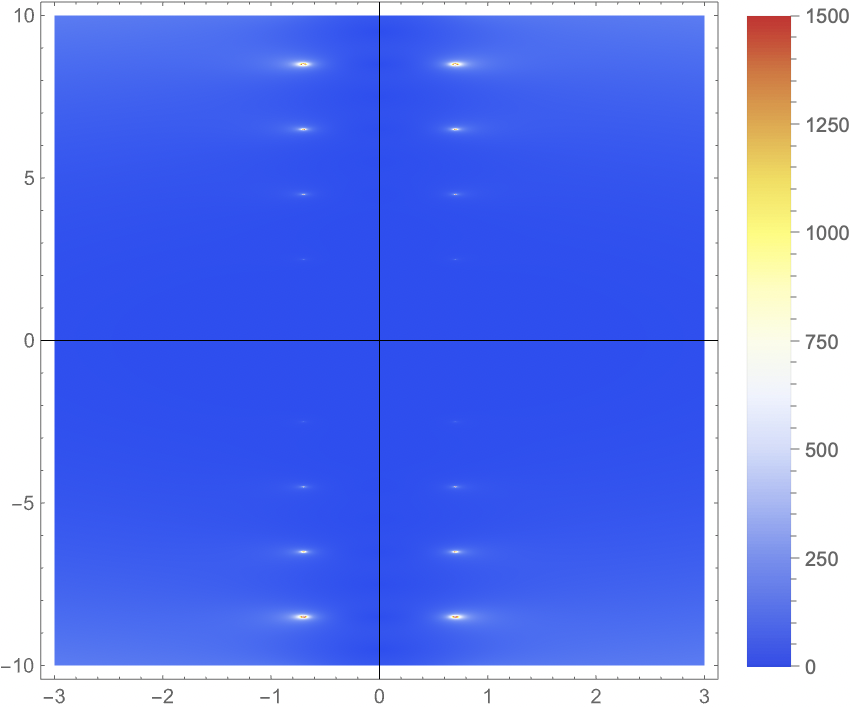}
            \caption[]%
            {{\small \centering  $\beta=2\pi$}}
            \label{fig:symds}
        \end{subfigure}
        \caption{$\left|\mathcal{G}^{\rm Sym}_{l=1}(\beta,\omega)\right|$ for a principal-series scalar with $\Delta = \frac{3}{2} + 0.7 i$ in $dS_4$ in complex $\omega$-plane. {The white spots mark the poles.}}
\end{figure}

At a generic inverse temperature $\beta$, \eqref{eq:wightmanretarded} has two kinds of poles. In addition to the resonance and anti-resonance poles inherited from the spectral function \eqref{eq:scalarspectral}, \eqref{eq:wightmanretarded} has poles at the Matsubara and anti-Matsubara frequencies associated with the inverse temperature $\beta$:
\begin{align}\label{eq:matfreq}
    i\omega^{\rm Mat}_{k} = \frac{2\pi k}{\beta} \; , \qquad k \in \mathbb{Z}\setminus \{0\} \; .
\end{align}
An example is shown in figure \ref{fig:symgeneric}.

\paragraph{When $\beta = 2\pi \mathbb{Q}$}

At the inverse temperatures
\begin{align}
    \beta = \frac{2\pi p}{q} \;,
\end{align}
where $p,q\in\mathbb{N}$ are co-prime integers, there are (partial) cancellations between the Matsubara poles \eqref{eq:matfreq} and the zeros \eqref{eq:spectralzeros}. Specifically, the would-be Matsubara poles at 
\begin{align}
    i\omega_k = kq  \; , \qquad k \in \mathbb{Z}\setminus \{0\}  \; ,
\end{align}
are absent. An example is shown in figure \ref{fig:symspecial}. For $p=1$, the Matsubara poles \eqref{eq:matfreq} are completely absent.

\paragraph{When $\beta = \beta_{\rm dS}= 2\pi$}

In this case of particular interest (see figure \ref{fig:symds}), \eqref{eq:wightmanretarded} becomes
\begin{align}\label{eq:symdS}
    \mathcal{G}^{\rm Sym}_l (2\pi,\omega) \propto
    \cosh \pi \omega \left| \Gamma\left( \frac{\Delta+l-i\omega}{2}\right) \Gamma\left( \frac{\bar\Delta+l-i\omega}{2}\right)  \right|^2 \; ,
\end{align}
where there is a one-to-one cancellation between the Matsubara poles \eqref{eq:matfreq} and the zeros \eqref{eq:spectralzeros}. 

\subsubsection{Two-sided function}

Before ending this section, we point out that we have been assuming that the 2-sided correlator {(cf. definition in \eqref{eq:2sided})} at the dS temperature $\beta =2\pi$, related to the spectral function by\footnote{The two-sided correlator has been studied extensively in other
contexts, most notably AdS/CFT. Its positivity and analytic properties in unitary and holographic theories are reviewed in \cite{Festuccia:2005pi,festuccia2007black,Dodelson:2023vrw}.}
\begin{align}\label{eq:scalar2sided}
    \mathcal{G}^{12}_l (\omega) = \frac{\mathcal{G}^C_l (\omega)}{2\sinh \pi \omega} \propto \Gamma\left( \frac{\Delta+l+i\omega}{2}\right) \Gamma\left( \frac{\bar\Delta+l+i\omega}{2}\right)\Gamma\left( \frac{\Delta+l-i\omega}{2}\right) \Gamma\left( \frac{\bar\Delta+l-i\omega}{2}\right) \;,
\end{align}
can be written as an absolute value squared and is thus real and positive. This is not true for general complex values of $\Delta$. However, this is indeed the case for $\Delta$ valued in the principal or complementary series, consistent with expectations from unitarity \cite{Dodelson:2023vrw}.

Another interesting comment is that using the Weierstrass form of the gamma function,
\begin{align}\label{eq:Weierstrass}
    \Gamma(z)=\frac{e^{-\gamma z}}{z}\prod_{n=1}^\infty \left(1+\frac{z}{n} \right)^{-1} e^{z/n} \; ,
\end{align}
we can recast \eqref{eq:scalar2sided} into the product formula 
\begin{align}\label{eq:thermalproduct}
    \mathcal{G}^{12}_l (\omega) = \frac{\mathcal{G}^{12}_l (\omega=0)}{\prod_{n=0}^{\infty}\left(1-\left(\frac{\omega}{\omega^\Delta_{nl}}\right)^2 \right)\left(1-\left(\frac{\omega}{\omega^{\bar\Delta}_{nl}}\right)^2 \right)} \; . 
\end{align}
A formula of this form was argued by \cite{Dodelson:2023vrw} for holographic thermal correlators in AdS/CFT.

\section{Static patch correlators from $S^{d+1}$}\label{sec:thermalfromsphere}

In this section, we explicitly compute the static patch correlators in the previous section and show how they can be obtained from analytically continued $S^{d+1}$ correlators.

\subsection{Worldline thermal correlators as resonance sums}

If one restricts the two points to be exactly on the worldline at $x=y=0$, $SO(d)$ invariance implies that only the $l=0$ mode can contribute to the full worldline correlators:  
\begin{align}
   G(t)= \frac{1}{{\rm Vol}(S^{d-1})}\int_{-\infty}^\infty d\omega \, e^{-i\omega t} \mathcal{G}_{l=0}(\omega) \;.
\end{align}
Here $G(t)$ can be any of the correlators discussed in the previous section. The factor $\frac{1}{{\rm Vol}(S^{d-1})}$ comes from the $l=0$ spherical harmonics $Y_{l=0}$ in the expansions of the full correlators in terms of the reduced ones, such as in \eqref{eq:retardrelation}. A general way to proceed is to close the contour on the lower (upper) half complex $\omega$-plane for $t>0$ ($t<0$), which then picks up contributions from poles of $\mathcal{G}_{l=0}(\omega)$. For example, the retarded function is a sum over resonances:
\begin{align}\label{eq:GRWL}
    G^R (t) 
    =&\frac{\Gamma\left(\frac{d}{2}-\Delta \right)}{\Gamma\left( \frac{d}{2}\right)\Gamma\left(1-\Delta \right)}\theta(t)\sum_{n=0}^\infty \frac{1}{n!} \frac{\left(\frac{d}{2}\right)_n \left(\Delta\right)_n}{\left(1-\frac{d}{2}+\Delta\right)_n}e^{-(\Delta+2n)t}+ \left(\Delta \leftrightarrow \bar \Delta\right)
    \nn\\
    =&\theta(t) \frac{\sin \left( \pi \Delta\right)\Gamma\left( \frac{d}{2}-\Delta\right) \Gamma\left(\Delta\right) }{ {\rm Vol}(S^{d+1})\Gamma\left( \frac{d}{2}+1\right)} e^{-\Delta t}  \,_2F_1\left( \Delta , \frac{d}{2} , 1 - \frac{d}{2} + \Delta, e^{-2t} \right) + \left(\Delta \leftrightarrow \bar \Delta\right) \; .
\end{align}
We have used ${\rm Vol}(S^{d+1})=\frac{2\pi}{d}{\rm Vol}(S^{d-1})$. Such a resonance/QNM sum was obtained in \cite{Jafferis:2013qia} for complementary-series scalars in $dS_4$, where a Fock space was built using the resonances normalized with respect to an `R-norm'. Our approach bypasses the introduction of the R-norm and treats both principal and complementary series uniformly. Similarly, the advanced function is a sum over anti-resonances.
For completeness, we also write down the spectral function:
\begin{align}\label{eq:GCWL}
     G^C (t) =&\frac{-i\sin \left( \pi \Delta\right)\Gamma\left( \frac{d}{2}-\Delta\right) \Gamma\left(\Delta\right) }{ {\rm Vol}(S^{d+1})\Gamma\left( \frac{d}{2}+1\right)} e^{-\Delta |t|}  \,_2F_1\left( \Delta , \frac{d}{2} , 1 - \frac{d}{2} + \Delta, e^{-2|t|} \right) + \left(\Delta \leftrightarrow \bar \Delta\right) \;. 
\end{align}
As emphasized earlier, these functions are insensitive to $\beta$. For the $\beta$-sensitive functions such as the symmetric Wightman function $G_\beta^{\rm Sym} (t)$, the contour could pick up Matsubara poles as discussed in the previous section. For generic $\beta$, $G_\beta^{\rm Sym} (t)$ consists of two sums, one over the (anti-)Matsubrara poles and the other one over the (anti-)resonance poles, for which we have not found a closed-form formula. For $\beta=2\pi$, the Matsubara sum is absent, and we have
\begin{align}\label{eq:GsymWL}
    G_{\beta=2\pi}^{\rm Sym} (t) = \frac{\cos \left( \pi \Delta\right)\Gamma\left( \frac{d}{2}-\Delta\right) \Gamma\left(\Delta\right) }{2 {\rm Vol}(S^{d+1})\Gamma\left( \frac{d}{2}+1\right)} e^{-\Delta |t|}  \,_2F_1\left( \Delta , \frac{d}{2} , 1 - \frac{d}{2} + \Delta, e^{-2|t|} \right) + \left(\Delta \leftrightarrow \bar \Delta\right) \; .
\end{align}
We note that \eqref{eq:GRWL}, \eqref{eq:GCWL}, and \eqref{eq:GsymWL} are valid for any dimensions $d\geq 2$.

\subsection{Analytic continuation from $S^{d+1}$}

We now show how these Lorentzian correlators can be obtained from suitable analytic continuation from the $S^{d+1}$ Green function,
\begin{align}\label{eq:EuclideanGreen}
G_E(P(\mathbf{x},\mathbf{y})) =\frac{\Gamma(\Delta)\Gamma(\bar{\Delta})}{\text{Vol} (S^{d+1})\Gamma(d+1)}\,_2F_1 \left(\Delta,\bar{\Delta};\frac{d+1}{2}; \frac{1+P(\mathbf{x},\mathbf{y})}{2}\right) \; , \qquad -1<P<1\; , 
\end{align}
where $\text{Vol} (S^{d+1})=\frac{2\pi^\frac{d+2}{2}}{\Gamma(\frac{d+2}{2})}$ is the volume for a unit round $S^{d+1}$, and $P(\mathbf{x},\mathbf{y})\equiv \cos \sigma (\mathbf{x},\mathbf{y})$ where $\sigma (\mathbf{x},\mathbf{y})$ the geodesic distance between $\mathbf{x},\mathbf{y}\in S^{d+1}$. The Green function \eqref{eq:EuclideanGreen} satisfies the Euclidean equation of motion with a delta-function source,
\begin{align}\label{eq:sphereGreeneom}
    \left(-\nabla_\mathbf{x}^2+m^2 \right)G_E(P(\mathbf{x},\mathbf{y})) = \frac{1}{\sqrt{g}}\delta^{(d+1)}(\mathbf{x},\mathbf{y}) \; ,
\end{align}
and can be represented as a path integral for the free scalar on $S^{d+1}$:
\begin{align}\label{eq:sphereGreen}
 	G_E(P(\mathbf{x},\mathbf{y}))  = \frac{\int \mathcal{D}\phi \,  \phi(\mathbf{x})\phi(\mathbf{y})\, e^{-S_E[\phi]}}{\int \mathcal{D}\phi \, e^{-S_E[\phi]}}\; , \qquad S_E = \frac{1}{2}\int_{S^{d+1}} \sqrt{g}\left[ \left(\partial \phi\right)^2+m^2 \phi^2 \right] \; .
\end{align}
Therefore, from the {\it global} dS perspective, \eqref{eq:EuclideanGreen} is the two point function for the free scalar in the Bunch-Davies vacuum where $\mathbf{x}$ and $\mathbf{y}$ are {\it spacelike}-separated. 

\begin{figure}[H]
    \centering
   \includegraphics[width=0.55\textwidth]{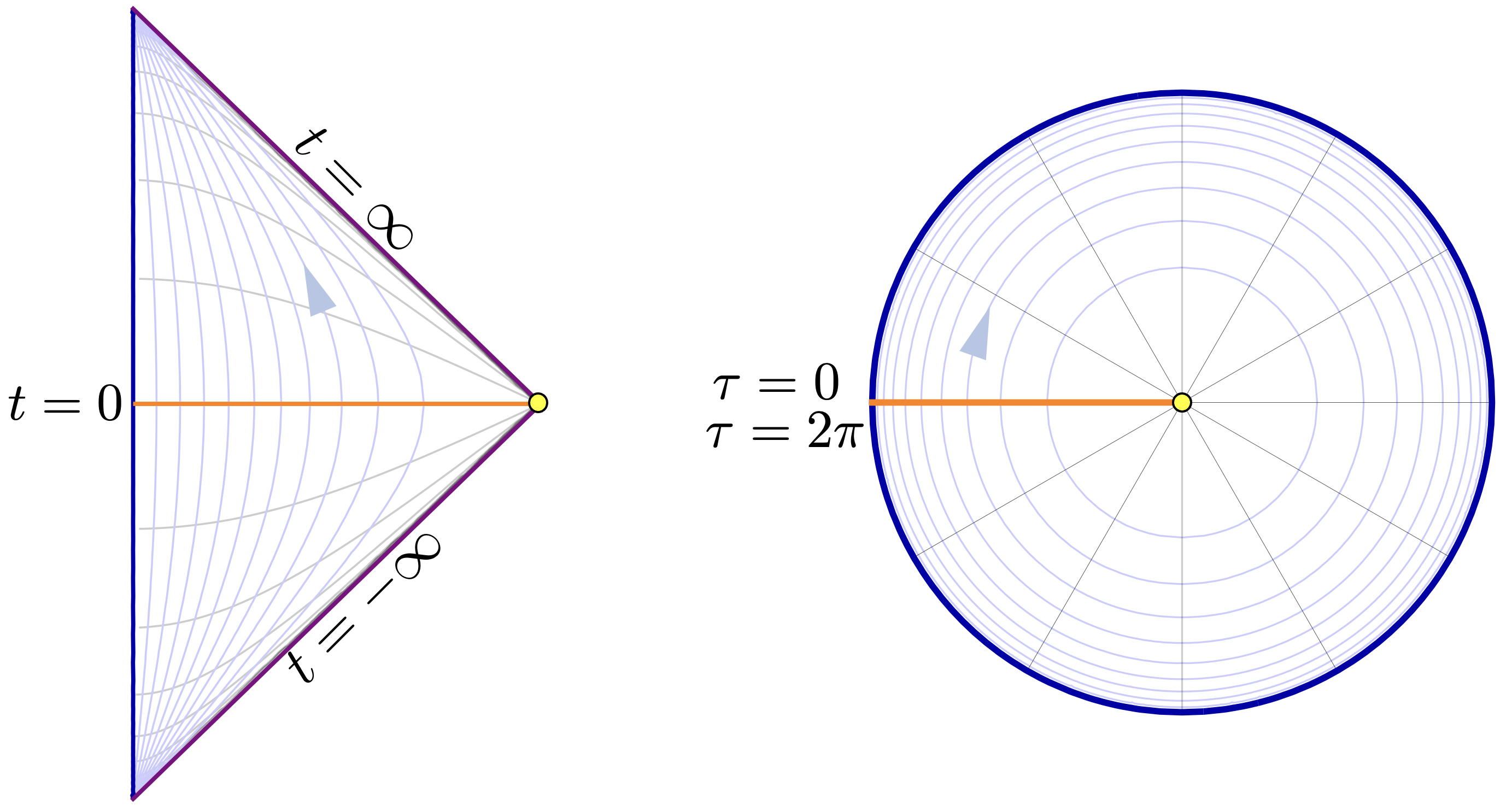}
        \caption{Left: Penrose diagram for a static patch. The yellow dot indicates the bifurcation surface at $r=\ell$. Right: Upon the Wick rotation $t \to -i\tau$, $\tau\sim\tau+2\pi$, the static patch becomes the sphere, with $\tau$ parametrizing the thermal circle.}
        \label{pic:staticpatch}
\end{figure}

From the static patch point of view, since $S^{d+1}$ can be obtained by Euclideanizing the static patch time $t\to -i\tau$ followed by the identification $\tau \sim \tau +2\pi$ (see figure \ref{pic:staticpatch}), the same quantity \eqref{eq:sphereGreen} can be interpreted as a thermal correlator in a static patch at inverse temperature $\beta_{\rm dS}=2\pi$, again with $\mathbf{x}$ and $\mathbf{y}$ spacelike-separated, i.e.  
\begin{align}
  \frac{\int \mathcal{D}\phi \,  \phi(\mathbf{x})\phi(\mathbf{y})\, e^{-S_E[\phi]}}{\int \mathcal{D}\phi \, e^{-S_E[\phi]}}= \frac{1}{Z(\beta_{\rm dS})}\Tr \, \left( e^{-\beta_{\rm dS} \hat H} \hat\phi(\mathbf{x})\hat\phi(\mathbf{y}) \right) \; , \quad Z(\beta) \equiv \Tr \, e^{-\beta \hat H} \; .
\end{align}
We recall that for spacelike separated points, the commutator-type functions are trivial, and all the Wightman-type functions are equivalent.

\paragraph{Analytic continuation to real time}

To continue to the region where $\mathbf{x}$ and $\mathbf{y}$ are {\it timelike}-separated, we need to Wick-rotate $\sigma\to it$, which takes $P$ to $P\in ( 1, \infty)$ where \eqref{eq:EuclideanGreen} has a branch cut. With an $i\epsilon$ prescription $P \rightarrow P\mp i\epsilon$ ($\epsilon>0$), we can safely perform the analytic continuation. After a linear transformation,
\begin{align} \label{eq:linearTrans}
		\,_2F_1(a,b,c,z) =& \frac{\Gamma(c)\Gamma(b-a)}{\Gamma(b) \Gamma(c-a)} (-z)^{-a} \,_2F_1\left(a,1-c+a;1-b+a;\frac{1}{z}\right) \nonumber \\
		&  + \frac{\Gamma(c)\Gamma(a-b)}{\Gamma(a)\Gamma(c-b)} (-z)^{-b} \,_2F_1\left(b,1-c+b;1-a+b,\frac{1}{z}\right)  \;,
\end{align}
(valid for $|\arg(-z)| < \pi$), and Kummer's quadratic  transformation,
\begin{equation} \label{eq:Kummer}
		\,_2F_1(a,b,a-b+1,z) = (1+\sqrt{z})^{-2a} \,_2F_1\left(a,a-b+\frac{1}{2},2a-2b+1,\frac{4 \sqrt{z}}{(1+\sqrt{z})^2}\right) \; ,
\end{equation}
we find\footnote{The $i \epsilon$ prescription determines whether the phase of $-z$ in the transformation \eqref{eq:linearTrans} is $+(\pi - \epsilon)$ or $-(\pi - \epsilon)$.}
\begin{align}\label{eq:continuedGE}
	G_E(P\mp i \epsilon) 
 =& \frac{e^{\mp i \pi \Delta} \Gamma(\Delta)\Gamma(\frac{d}{2}-\Delta) }{2{\rm Vol}(S^{d+1})\Gamma\left( \frac{d}{2}+1\right)}e^{-\Delta |t| }\, _2F_1\left( \Delta, \frac{d}{2}, 1 - \frac{d}{2} + \Delta, e^{-2|t|} \right) + (\Delta \leftrightarrow \bar \Delta) \; .
\end{align}
It is now clear how the Lorentzian $dS_{d+1}$ correlators are obtained from those on $S^{d+1}$. In particular,
\begin{align}\label{eq:massanacno}
    G^C (t) &= G_E(P- i \epsilon) - G_E(P+i \epsilon) \; , \qquad 
    G_{\beta = 2\pi}^{\rm Sym} (t) = \frac{G_E(P- i \epsilon) + G_E(P+i \epsilon)}{2}\;. 
\end{align}


\section{$SO(1,d+1)$ Harish-Chandra character: a static patch viewpoint} \label{sec:characterStaticPatchViewpoint}

Particles and fields in $dS_{d+1}$ are naturally classified according to unitary irreducible representations (UIRs) of the de Sitter group $SO(1,d+1)$ \cite{Basile:2016aen}. A $SO(1,d+1)$ UIR $\mathcal{V}_{[\Delta, \mathbf{s}]}$ is labeled by an $SO(1,1)$ weight $\Delta$ and an $SO(d)$ weight $\mathbf{s}$, and can be realized as the single-particle Hilbert space in global quantization \cite{Sun:2021thf}. A useful object that encodes the information of a UIR is its Harish-Chandra character, defined as a trace over the representation space,
\begin{align}
    \chi (g) = \tr_{[\Delta, \mathbf{s}]} \, R(g) \;, \qquad g\in SO(1,d+1) \; ,
\end{align}
viewed as a distribution to be integrated against smooth test functions on $SO(1,d+1)$. The classifications of $SO(1,d+1)$ UIRs and their
characters were studied in \cite{10.3792/pja/1195522333,10.3792/pja/1195523378,10.3792/pja/1195523460} (see \cite{Sun:2021thf} for a review). While it is not difficult to consider the character for a general $SO(1,d+1)$ element, we will focus on the element $g=e^{-i\hat H t}$ which generates the $SO(1,1)$ subgroup. $\hat H$ acts on the de Sitter geometry globally as a boost and on a static patch as the Hamiltonian. The character for a scalar with a mass squared $m^2 = \Delta (d-\Delta)>0$ reads explicitly
\begin{align}\label{eq:HCcharacter}
    \chi(t) 
    = \frac{e^{-\Delta t}+e^{-\bar \Delta t}}{|1-e^{-t}|^d} \; .
\end{align}
Note that this satisfies the property $\chi(-t)=\chi(t)$. 

There are two key related observations in \cite{Anninos:2020hfj}. First, the characters admit an expansion in terms of the resonance poles. For scalars, it is evident by expanding \eqref{eq:HCcharacter} in powers of $e^{-|t|}$:
\begin{align}
    \chi (t) = \sum_{n=0}^\infty \sum_{l=0}^\infty D^d_l \, \left( e^{-i \omega^\Delta_{nl} |t|}+e^{-i \omega^{\bar\Delta}_{nl} |t|}\right)\; ,
\end{align}
where $\omega^\Delta_{nl}$ and $\omega^{\bar\Delta}_{nl}$ are the resonance frequencies \eqref{eq:resonancepoles}. Characters for more general UIRs also admit such an expansion \cite{Sun:2020sgn}. The second observation is that the Fourier transform of the character,
\begin{align}\label{eq:doschar}
    \tilde\rho^\text{dS}(\omega)\equiv \int_{-\infty}^\infty \frac{dt}{2\pi}e^{i\omega t} \chi(t) \;,
\end{align}
defines a spectral density for the single-particle Hamiltonian. With \eqref{eq:doschar}, the Lorentzian calculation of the canonical thermal partition function for a scalar on the static patch can be rendered well-defined, and exactly reproduces the Euclidean path integrals on $S^{d+1}$.


\subsection{Scattering matrices and relative spectral density}\label{sec:Smatrix}

In \cite{Law:2022zdq}, the precise physical meaning of the spectral density \eqref{eq:doschar} was clarified and generalized to other static spherically symmetric black holes. We now review the key ideas specializing to the case of a dS static patch, which will set the stage for our main result in section \ref{sec:specchar}.

\paragraph{Resonance poles vs scattering poles}

For a fixed $l\geq 0$, the ratio between the outgoing and ingoing coefficients for the normalizable mode \eqref{near hor} defines a scattering phase,
\begin{align}\label{eq:Smatrixdef}
    \mathcal{S}_l (\omega)  \equiv \frac{B_{\omega l}}{B_{-\omega l}}\;,
\end{align}
which maps an incoming wave from the past horizon to an outgoing wave to the future horizon:
\begin{figure}[H]
    \centering
   \includegraphics[width=0.65\textwidth]{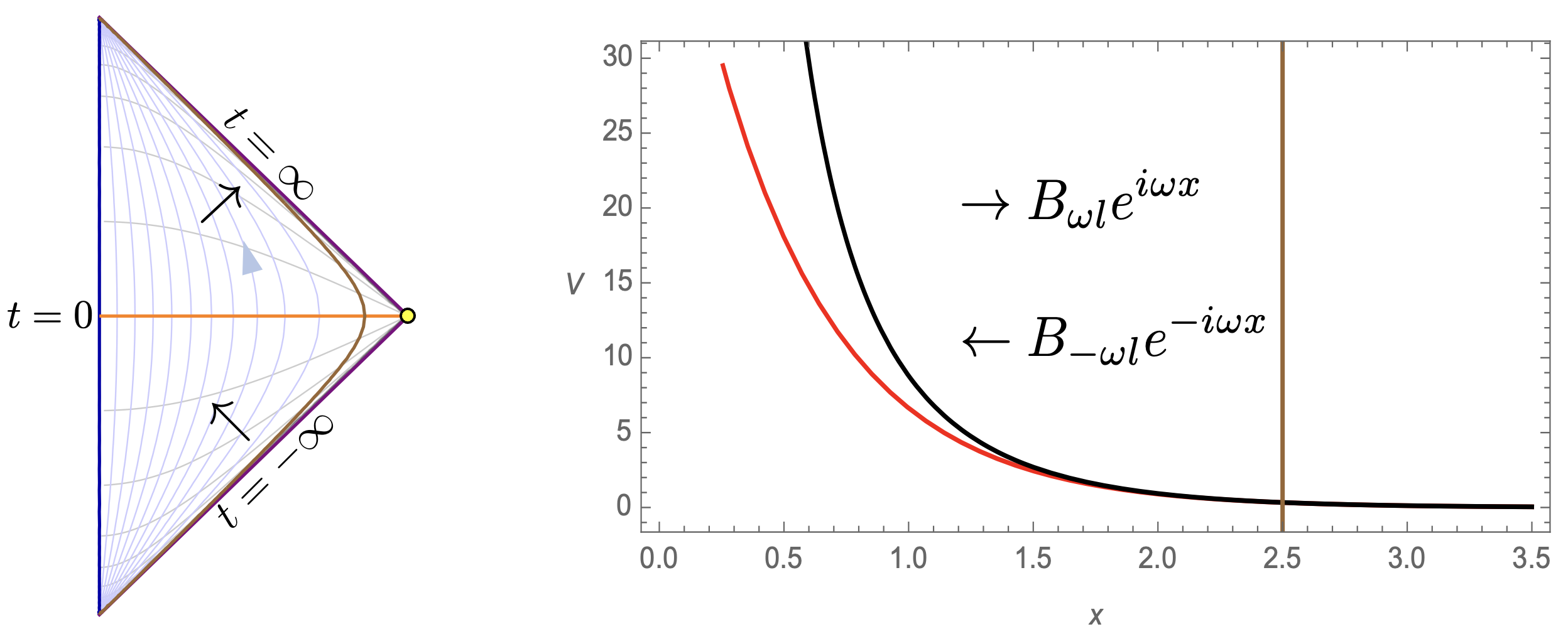}
        \caption{Left: The Penrose diagram for a static patch, with the black arrows indicating incoming and outgoing waves. Right: The scattering potential \eqref{eq:potential} for a principal-series scalar with $\Delta=\frac{3}{2}+0.1 i$ and $l=3$ in $dS_4$, which is hardly distinguishable from the Rindler potential (red) in \eqref{appeq:rindlerscatt} as $x\to \infty$. In these diagrams, the brown lines indicate a brick wall regulator \cite{tHooft:1984kcu} at $x=R=2.5$.}
        \label{pic:scattering}
\end{figure}
The poles/zeros of \eqref{eq:Smatrixdef}, which we call the {\bf scattering poles/zeros}, correspond to modes that are regular at the observer's location and purely outgoing/incoming at the horizon. Explicitly, \eqref{eq:Smatrixdef} can be put into the product form\footnote{The factorization into a geometry-dependent piece and a universal Rindler factor is not special to the dS static patch, but holds quite generally for static black holes with nonextremal horizons \cite{Law:2022zdq,Grewal:2022hlo}, arising from the universal near-horizon Rindler structure of the wave equation.}
\begin{gather}\label{eq:dSSmat}
    \mathcal{S}_l (\omega)  = \mathcal{S}_l^\text{dS} (\omega) \mathcal{S}^\text{Rin} \left(\omega\right)\;,\nn\\
    \mathcal{S}_l^\text{dS} (\omega)\equiv \frac{\Gamma\left(\frac{\Delta+l- i  \omega}{2}\right) \Gamma\left(\frac{\bar\Delta+l- i  \omega}{2}\right)}{\Gamma\left(\frac{\Delta+l+i  \omega}{2}\right) \Gamma\left(\frac{\bar\Delta+l+ i \omega}{2}\right)}  \; , \qquad 
    \mathcal{S}^\text{Rin} \left(\omega\right) = 2^{-2i\omega} \frac{\Gamma(i\omega)}{\Gamma(-i\omega)} \; .
\end{gather}
The poles (zeros) of $\mathcal{S}_l^\text{dS} (\omega)$ capture the (anti-)resonance poles \eqref{eq:resonancepoles}. Meanwhile, as pointed out in \cite{Law:2022zdq} and reviewed in appendix \ref{app:Rin2D}, $\mathcal{S}^\text{Rin} \left(\omega\right)$ is exactly the scattering phase \eqref{appeq:RindlerS} associated with the wave equation \eqref{appeq:rindlerscatt} on a 2D Rindler space with $\beta=\beta_{\rm dS}=2\pi$ and $a=\log 2$. The poles (zeros) of $\mathcal{S}^\text{Rin} \left(\omega\right)$, 
\begin{align}\label{eq:dSmatfreq}
    i\omega^{\rm Mat}_{k} = \pm k \; , \qquad k \in \mathbb{N} \; ,
\end{align}
correspond to the (anti-)Matsubara frequencies associated with the dS temperature. 

We therefore have an interesting conclusion that (anti-)resonance poles do not coincide with scattering poles (zeros): there exist modes that satisfy the (anti-)QNM boundary conditions yet are not poles of the retarded (advanced) Green functions.

\paragraph{Scattering phases and spectral density}

While the density of states for a system with a continuous spectrum diverges, its {\it difference} relative to another system with a continuous spectrum is well-defined. The {\it relative} spectral density is captured by the (generalized) Krein-Friedel-Lloyd formula
\begin{align}\label{eq:Krein}
    \Delta\rho_l(\omega)=\frac{1}{2\pi i}\partial_\omega \left(\log \mathcal{S}_l(\omega) - \log \bar{\mathcal{S}} (\omega)\right) \; .
\end{align}
Here $\bar{\mathcal{S}} (\omega)$ is the scattering phase for the reference problem. An observation in \cite{Law:2022zdq} is that if we take the reference problem to be the one associated with a wave equation on the 2D Rindler space with $\beta=\beta_{\rm dS}=2\pi$, i.e. we take $\bar{\mathcal{S}} (\omega)= \mathcal{S}^\text{Rin} \left(\omega\right)$, the resulting spectral density 
\begin{align}\label{eq:eq:dSspecden}
    \Delta\rho^\text{dS}_l(\omega)=\frac{1}{2\pi i}\partial_\omega \log \mathcal{S}_l^\text{dS} (\omega)
\end{align}
coincides with the one defined by the Fourier-transform of the Harish-Chandra character \eqref{eq:doschar} upon summing over $l\geq 0$ (up to an infinite $\omega$-independent constant), i.e.
\begin{align}\label{eq:rhocoincide}
    \tilde\rho^\text{dS}(\omega)=\sum_{l=0}^\infty D_l^d \Delta\rho^\text{dS}_l(\omega) \; .
\end{align}

\subsection{Relation with the spectral function}\label{sec:specchar}

We are now ready to accomplish the main goal of this paper: establishing a direct connection between the Harish-Chandra character \eqref{eq:HCcharacter} and the correlators studied in detail in section \ref{sec:staticLor}. Specifically, the spectral function
\begin{align}\label{eq:quantumGC}
    G^C(t,x,\Omega|0,y,\Omega')&\equiv \bra{0} \left[\hat \phi (t,x,\Omega),\hat \phi (0,y,\Omega')\right] \ket{0}\nn\\
    &=-i \left(G^R(t,x,\Omega|0,y,\Omega')-G^A(t,x,\Omega|0,y,\Omega') \right)
\end{align}
plays a distinguished role. Similar to \eqref{eq:retardrelation}, we can decompose \eqref{eq:quantumGC} as a sum over $l\geq 0$:
\begin{align}\label{eq:GCrelation}
    G^C(t;x,\Omega| y,\Omega')=\tanh^{\frac{1-d}{2}}x \,\tanh^{\frac{1-d}{2}}y \, \sum_{l=0}^\infty \bar G^C_l(t;x| y) \, Y_l (\Omega)\, Y_l (\Omega') \;.
\end{align}
Our analysis starts by deriving a formula for the Fourier transform
\begin{align}\label{eq:Gcdef}
    \bar{\mathcal{G}}_l^C(\omega;x|y)= -i\left(\bar{\mathcal{G}}^R_l(\omega;x|y)-\bar{\mathcal{G}}_l^A(\omega;x|y)\right) \;,
\end{align}
by noting that both $\bar{\mathcal{G}}^R_l(\omega;x|y)$ and $\bar{\mathcal{G}}_l^A(\omega;x|y)$ satisfy \eqref{eq:rescaledGeomFourier}, implying
\begin{align}
    \bar{\mathcal{G}}^R_l(\omega;x|y)-\bar{\mathcal{G}}_l^A(\omega;x|y)
    =& \int_0^\infty dz \left[\bar{\mathcal{G}}_l^A(\omega;y|z)\partial^2_z \bar{\mathcal{G}}^R_l(\omega;x|z) -\bar{\mathcal{G}}^R_l(\omega;x|z)\partial^2_z \bar{\mathcal{G}}_l^A(\omega;y|z) \right] \nn\\
    =& \left[\bar{\mathcal{G}}_l^A(\omega;y|z)\partial_z \bar{\mathcal{G}}^R_l(\omega;x|z) -\bar{\mathcal{G}}^R_l(\omega;x|z)\partial_z \bar{\mathcal{G}}_l^A(\omega;y|z)  \right]^{z=\infty}_{z=0} \;. 
\end{align}
The last line can then be computed using the near-origin behaviors \eqref{eq:nearorigin} and near-horizon asymptotics \eqref{eq:Jostasym} of the Jost solution, resulting in
\begin{align}\label{eq:spectralfnformula}
    \boxed{\bar{\mathcal{G}}_l^C(\omega;x|y)=\frac{\omega}{\pi} \frac{{\bar\psi}_{\omega l}^{\rm n.}(x){\bar\psi}_{\omega l}^{\rm n.}(y) }{W^{\rm Jost}_l(\omega)W^{\rm Jost}_l(-\omega)}} \;. 
\end{align}
Notice that since the overall normalizations of $\bar{\mathcal{G}}_l^R$ and $\bar{\mathcal{G}}_l^A$ are uniquely fixed by \eqref{eq:rescaledGeomFourier}, that of $\bar{\mathcal{G}}_l^C$ is uniquely fixed as well.

\paragraph{The spectral trace}

Regarding \eqref{eq:Gcdef} as the integration kernel associated with an integral operator $\hat{\mathcal{G}}^C_l (\omega)$, it is natural to consider its trace as an integral over the diagonal element
\begin{align}
    ``\tr" \hat{\mathcal{G}}^C_l (\omega) = \int_0^\infty dx \, \bar{\mathcal{G}}_l^C(\omega;x|x)=\frac{\omega}{\pi}  \int_0^\infty dx\frac{{\bar\psi}_{\omega l}^{\rm n.}(x){\bar\psi}_{\omega l}^{\rm n.}(x) }{W^{\rm Jost}_l(\omega)W^{\rm Jost}_l(-\omega)} \;, 
\end{align}
which however is ill-defined because the integral diverges in the region $x\to \infty$. We cut off the integral at $x=R$ (see figure \ref{pic:scattering}) and study its large $R$ asymptotics by the following trick. From \eqref{eq:scalareomsep}, one can deduce 
\begin{align}
	  (a^2-\omega^2) {\bar\psi}_{\omega l}^\text{n.}  (x)  {\bar\psi}_{a l}^\text{n.}  (x)={\bar\psi}_{a l}^\text{n.}  (x)\partial^2_x{\bar\psi}_{\omega l}^\text{n.}  (x)-{\bar\psi}_{\omega l}^\text{n.}  (x) \partial^2_x  {\bar\psi}_{a l}^\text{n.}  (x) \; .
\end{align}
Taking the derivative of both sides with respect to $a$ and setting $a=\omega$ implies the formula
\begin{align}
	\int_0^R dx \, {\bar\psi}_{\omega l}^\text{n.}  (x) {\bar\psi}_{\omega l}^\text{n.}  (x)  =& \frac{1}{2\omega}\int_0^R dx \left( \partial_\omega{\bar\psi}_{\omega l}^\text{n.}  (x)\partial^2_x{\bar\psi}_{\omega l}^\text{n.}  (x)-{\bar\psi}_{\omega l}^\text{n.}  (x) \partial_\omega \partial^2_x  {\bar\psi}_{\omega l}^\text{n.}  (x)\right)\nn\\
	=&\frac{1}{2\omega}	\left[ \partial_\omega{\bar\psi}_{\omega l}^\text{n.}  (x)\partial_x{\bar\psi}_{\omega l}^\text{n.}  (x)- {\bar\psi}_{\omega l}^\text{n.}  (x) \partial_\omega \partial_x  {\bar\psi}_{\omega l}^\text{n.}  (x) \right]_{x=0}^{x=R} \; .
\end{align}
Now, using \eqref{eq:nearorigin} and \eqref{near hor} results in
\begin{align}\label{eq:dSspectrace}
    \int_0^R dx \, \bar{\mathcal{G}}_l^C(\omega;x|x) = \frac{R}{2\pi \omega }+\frac{1}{4\pi i\omega}\partial_\omega \log \mathcal{S}_l (\omega)+O\left(\frac{1}{R}\right) \;. 
\end{align}
Here $\mathcal{S}_l (\omega)$ is the scattering phase \eqref{eq:Smatrixdef}. Motivated by the reasoning in \cite{Law:2022zdq} leading to considering a relative spectral density \eqref{eq:Krein}, we consider an analogous object for a reference scattering problem. In particular, if we pick the reference problem to be the Rindler scattering problem \eqref{appeq:rindlerscatt} with $\beta=2\pi$ and $a=\log 2$, we would have
\begin{align}\label{eq:Rinspectrace}
    \int_{-\infty}^R dx \, \bar{\mathcal{G}}_l^{{\rm Rin},C}(\omega;x|x) = \frac{R}{2\pi \omega }+\frac{1}{4\pi i\omega}\partial_\omega \log \mathcal{S}^{\rm Rin} (\omega)+O\left(\frac{1}{R}\right) \;. 
\end{align}
Note that the lower limit for the $x$-integral is $-\infty$. Now, taking the difference between \eqref{eq:dSspectrace} and \eqref{eq:Rinspectrace}, we can safely send $R\to\infty$. Comparing the result with \eqref{eq:eq:dSspecden} then establishes the relation
\begin{align}\label{eq:DOSspecfn}
    \boxed{\Delta \rho^\text{dS}_l(\omega)  =2 \, \omega\, \tilde \tr  \, \hat{\mathcal{G}}^C_l (\omega) } \;. 
\end{align}
Here $\tilde \tr $ denotes the renormalized trace:
\begin{align}\label{eq:renormalizedtrdef}
    \tilde \tr  \, \hat{\mathcal{G}}^C_l (\omega)  \equiv \lim_{R\to \infty} \left(\int_0^R dx \, \bar{\mathcal{G}}_l^C(\omega;x|x)-\int_{-\infty}^R dx \, \bar{\mathcal{G}}_l^{{\rm Rin},C}(\omega;x|x)\right) \;. 
\end{align}

\paragraph{Harish-Chandra character and the spectral function}

Now, we take the inverse Fourier transform of \eqref{eq:DOSspecfn} and sum over $l\geq 0$. The left hand side becomes the Harish-Chandra character $\chi(t)$. The first term in \eqref{eq:renormalizedtrdef} reproduces the full spectral function \eqref{eq:quantumGC} on the static patch \eqref{eq:tortoisemetric}. The second term \eqref{eq:renormalizedtrdef} is naturally related to the spectral function 
\begin{align}
    G^{\text{Rin},C}(t,x,\Omega|0,y,\Omega')&\equiv \bra{0} \left[\hat \phi (t,x,\Omega),\hat \phi (0,y,\Omega')\right] \ket{0}
\end{align}
for the scalar field on the near-horizon Rindler-like geometry \eqref{eq:tortoisemetricnearhor}. Strictly speaking, for each fixed $l$, the reduced radial scattering problem inherited from the Klein-Gordon equation is not exactly the Rindler problem \eqref{appeq:rindlerscatt} with $\beta=2\pi$ and $a=\log 2$; instead the parameter $a$ is shifted to an $l$-dependent value $a_l$. However, this simply translates to a finite shift $R\to R-a_l+\log 2$ in \eqref{eq:Rinspectrace}, which upon the inverse Fourier transform contributes only a contact term at $t=0$. Up to such contact terms, we conclude that the Harish-Chandra character is proportional to the time derivative of the {\it traced} spectral function: 
\begin{align}\label{eq:charGC}
    \boxed{\chi(t)  = 2i \frac{d}{dt} \tilde\tr \,  \hat{G}^C(t)} \;. 
\end{align}
Here the renormalized trace for the full spectral function is defined as
\begin{align}
   \tilde\tr \,\hat{G}^C(t) \equiv \lim_{R\to \infty} \left(\int_{\Sigma_R} \sqrt{-g^{tt}}\, G^C(t;x,\Omega| x,\Omega)-\int_{\Sigma^{\rm Rin}_R} \sqrt{-g_{\rm Rin}^{tt}}\, G^{{\rm Rin},C}(t;x,\Omega| x,\Omega) \right)\;. 
\end{align}
In this expression, the integrals are over the spatial slices of the static patch \eqref{eq:tortoisemetric} and its near-horizon Rindler-like geometry \eqref{eq:tortoisemetricnearhor} with a cutoff at $x=R$:
\begin{align}
    \int_{\Sigma_R} \equiv \int_0^R dx \int_{S^{d-1}} d\Omega \, \sqrt{g_{xx}} \,\tanh^{d-1}x \;, \qquad \int_{\Sigma^{\rm Rin}_R} \equiv \int_{-\infty}^R dx \int_{S^{d-1}} d\Omega \, \sqrt{g^{\rm Rin}_{xx}}\;. 
\end{align}
The relation \eqref{eq:charGC} is our central result.  It expresses the Harish-Chandra character---originally defined purely within $SO(1,d+1)$ representation theory---in terms of a physical observable of the static patch that remains well-defined even in interacting theories. {It would be interesting to revisit the group-theoretic derivation of the Harish–Chandra character, which might clarify the mathematical or physical origin of the matching \eqref{eq:charGC}.}

\subsubsection{A single-particle calculation of the character}

Naively, one might interpret \eqref{eq:HCcharacter} as a trace over the single-particle Hilbert space in the static patch and wish to compute it in the basis $\ket{\omega l}$ that diagonalizes the static patch Hamiltonian $\hat H$ and $SO(d)$ angular momentum $\mathbf{J}$. This leads to a non-sensible result \cite{Anninos:2020hfj}
\begin{align}
    \chi(t) \stackrel{?}{=} \int d\omega \sum_l \bra{\omega l} e^{-i \hat H t} \ket{\omega l} = 2\pi \sum_l \delta (0)\delta (t)\; . 
\end{align}
As explained in \cite{Anninos:2020hfj,Sun:2021thf}, one way to recover  \eqref{eq:HCcharacter} as some single-particle trace is to compute it in a basis on which all $SO(1,d+1)$ generators (in particular the $SO(1,1)$ generator) act regularly -- one labeled by the {\it global} $SO(d+1)$ angular momentum, for example. However, the static patch interpretation for these calculations is not completely transparent.

Our discussion leading to \eqref{eq:charGC} suggests an elementary single-particle calculation for the character \eqref{eq:HCcharacter} from the static patch point of view. The starting point is the matrix element in the position basis:
\begin{align}
    \bra{x,\Omega}e^{-i \hat H t}\ket{y,\Omega'} =  \sum_{l=0}^\infty \int_0^\infty d\omega \, e^{-i \omega t}\braket{x,\Omega|\omega,l}\braket{\omega,l| y,\Omega'} \;. 
\end{align}
Here the wave function 
\begin{align}\label{eq:wavefn}
    \braket{x,\Omega|\omega,l} = \sqrt{2\omega}|C_{\omega l}|\, \psi_{\omega l}^{\rm n.}(x)Y_l (\Omega)\; , \qquad \omega >0 \; , 
\end{align}
is $\delta$-function-normalized with respect to the canonical spatial norm:
\begin{align}\label{eq:wavefnnorm}
    \braket{\omega',l'|\omega,l} = \int_0^\infty dx \tanh^{d-1}x \int_{S^{d-1}}d\Omega \, \braket{\omega',l'|x,\Omega} \braket{x,\Omega|\omega,l} = \delta(\omega-\omega')\delta_{l l'}\;. 
\end{align}
Note that \eqref{eq:wavefnnorm} is the same as the normalization condition \eqref{eq:KGcoecondition} for $\omega,\omega'>0$, so the coefficient $C_{\omega l}$ in \eqref{eq:wavefn} coincides (up to a constant phase) with the coefficient \eqref{eq:scalarlincom} for the normalizable mode.

A trace-like object is given by the integral of the diagonal element:
\begin{align}\label{eq:worldlineGRtr}
    ``\tr" e^{-i \hat H t} =& \int_0^\infty dx \int_{S^{d-1}}d\Omega \bra{x,\Omega}e^{-i \hat H t}\ket{x,\Omega} \nn\\
    =&\sum_{l=0}^\infty \int_0^\infty  d\omega \, e^{-i \omega t} \int_0^\infty dx  \, \left|\braket{x,\Omega|\omega,l}\right|^2 \; .
\end{align}
Again, this trace is ill-defined because the $x$-integral on the right-hand side diverges in the region $x\to \infty$. By now, we know very well how to proceed: cut off the integral at $x=R$, subtract off the analogous integral in a reference Rindler scattering problem, and send $R\to \infty$ at the end. The calculation proceeds in the same way as the one leading to \eqref{eq:DOSspecfn}. The result is that
\begin{align}\label{eq:dosintegral}
    \Delta\rho_l^\text{dS}(\omega)= \lim_{R\to \infty} \left( \int_0^R dx \left|\braket{x,\Omega|\omega,l}\right|^2- \int_{-\infty}^R dx\, \left|\braket{x,\Omega|\omega,l}\right|_{\rm Rin}^2 \right) \;,
\end{align}
and therefore summing over $l$ we have
\begin{align}\label{eq:renomchart}
    \tilde\tr\, e^{-i \hat H t} 
    =&\sum_{l=0}^\infty D_l^d\int_0^\infty  d\omega \, e^{-i \omega t}  \Delta\rho^\text{dS}_l(\omega)\; .
\end{align}
Here the renormalized trace $\tilde\tr$ is defined analogously as in \eqref{eq:renormalizedtrdef}. To finally make a connection with the character, we invert \eqref{eq:doschar} to find
\begin{align}
    \chi(t)=&\sum_{l=0}^\infty D_l^d\int_{-\infty}^\infty  d\omega \, e^{-i \omega t}  \Delta\rho^\text{dS}_l(\omega)\; .
\end{align}
Notice that the integral over $0<\omega<\infty$ gives \eqref{eq:renomchart}, while the integral over $-\infty<\omega<0$ gives the same quantity but with $t\to -t$. To summarize, the character is related to the elementary (renormalized) traces through
\begin{align}
    \boxed{\chi(t) = \tilde\tr \left( e^{-i \hat H t}+e^{i \hat H t}\right)} \;. 
\end{align}

\subsection{Revisiting principal and complementary characters}\label{sec:charpole}

In this section, we take a closer look at the analytic structure of the relative spectral density \eqref{eq:doschar}. By applying \eqref{eq:Weierstrass} in \eqref{eq:eq:dSspecden} (ignoring an infinite $\omega$-independent constant), one can separate the contributions from the resonance and anti-resonance poles as 
\begin{align}\label{eq:dossplit}
    \tilde\rho^\text{dS}(\omega)=\tilde \rho^R(\omega)+\tilde \rho^A(\omega) \; ,
\end{align}
where
\begin{align} \label{eq:dosRetarded}
    \tilde \rho^R(\omega) =& \tilde \rho^A(-\omega) = \sum_{l=0}^{\infty}\Delta \rho^R_l(\omega)= \frac{1}{2\pi i} \sum_{l=0}^{\infty}\sum_{n=0}^{\infty}D^d_l \left(\frac{1}{\omega^\Delta_{nl}-\omega} +\frac{1}{\omega^{\bar\Delta}_{nl}-\omega} \right) \;. 
\end{align}
Here $\omega^\Delta_{nl}$ and $\omega^{\bar\Delta}_{nl}$ are the resonance frequencies \eqref{eq:resonancepoles}.

\paragraph{Structure of the character}

To obtain the character, we perform the inverse Fourier transform
\begin{align}\label{eq:chiinverse}
    \chi (t) = \int_{-\infty}^\infty d
    \omega \, e^{-i\omega t}\tilde\rho^\text{dS}(\omega)\; . 
\end{align}
For both principal ($\Delta=\frac{d}{2}+i \nu \, , \nu\in\mathbb{R}$) and complementary ($0<\Delta<d$) series, the (anti-)resonance frequencies \eqref{eq:resonancepoles} all lie in the lower (upper) half of the complex $\omega$-plane. 
\begin{figure}[H]
    \centering
   \begin{subfigure}{0.3\textwidth}
            \centering
            \includegraphics[height=4.0cm]{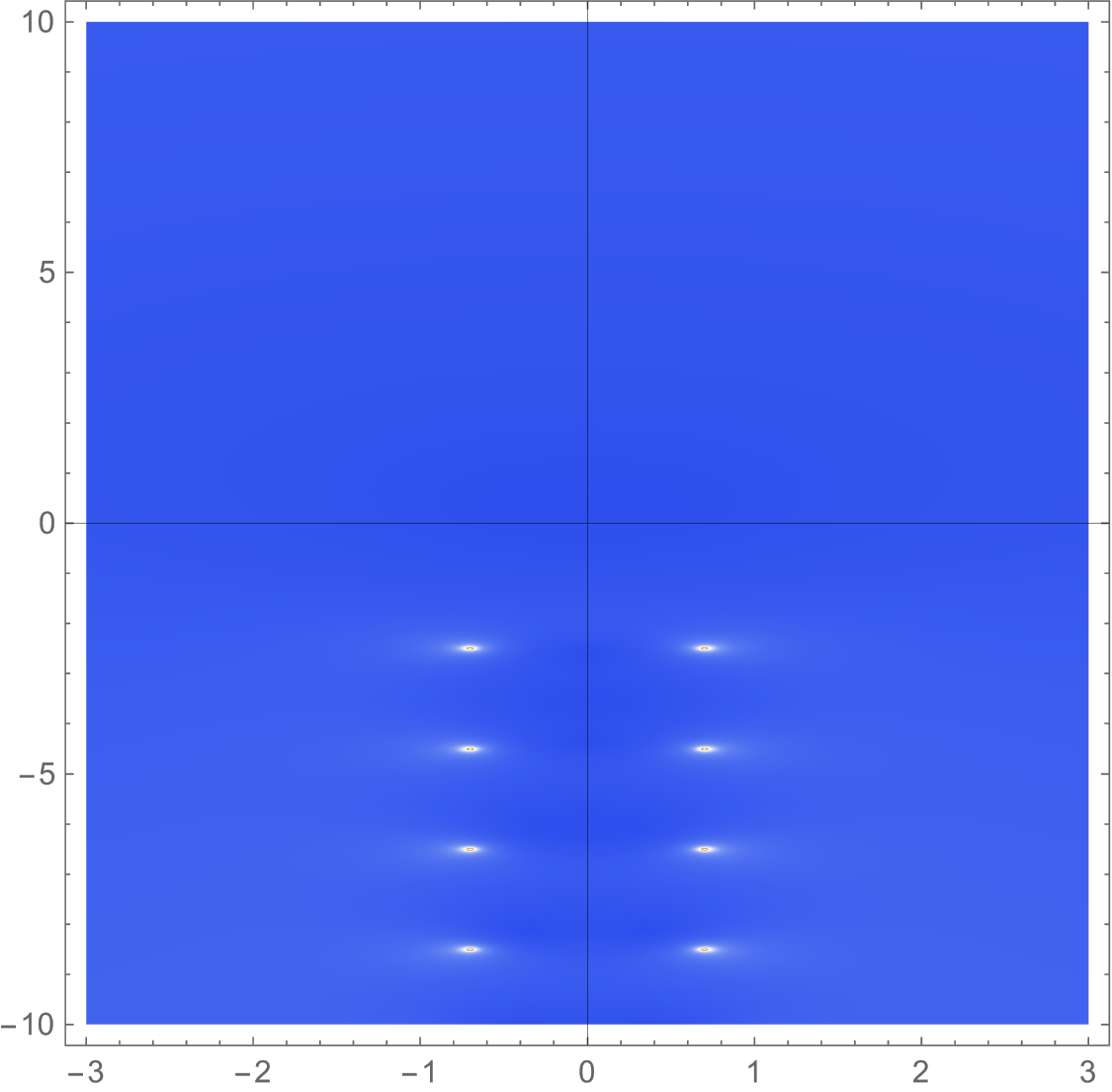}
            \caption[]%
            {{$\left|\Delta \rho^R_{l=1}(\omega)\right|$}}    
        \end{subfigure}
        \hspace{0.3cm}
        \begin{subfigure}{0.3\textwidth}  
            \centering 
            \includegraphics[height=4.0cm]{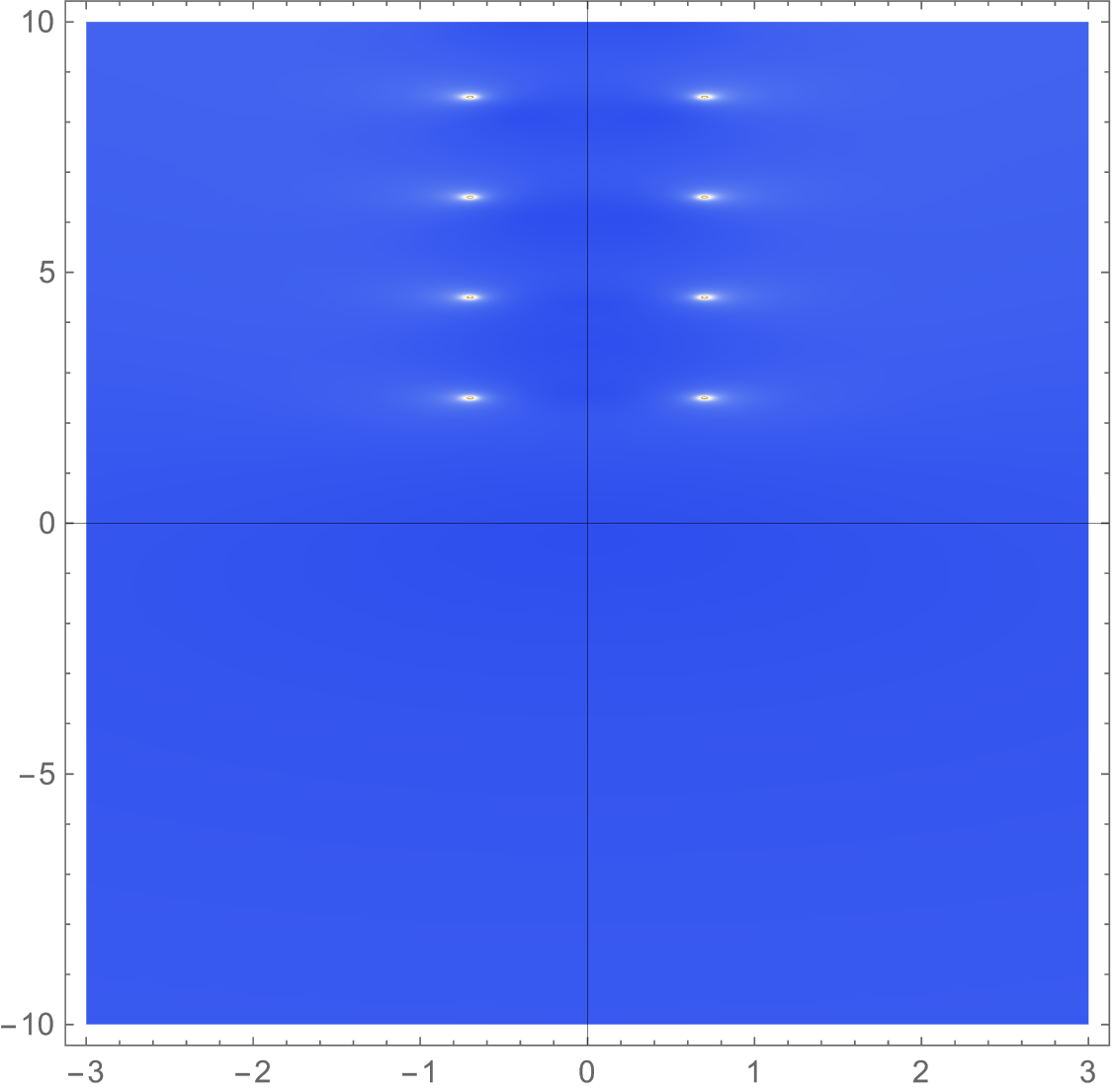}
            \caption[]%
            {{$\left|\Delta \rho^A_{l=1}(\omega)\right|$}}    
        \end{subfigure}
        \hspace{0.3cm}
        \begin{subfigure}{0.3\textwidth}  
            \centering 
            \includegraphics[height=4.0cm]{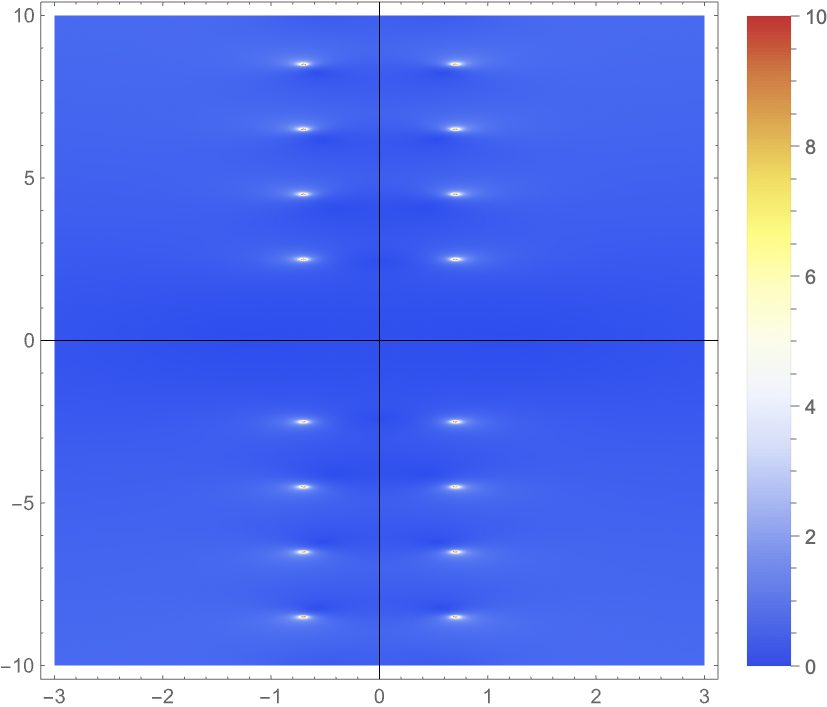}
            \caption[]%
            {{\small \centering  $\left|\Delta \rho_{l=1}(\omega)\right|$}}    
        \end{subfigure}
        \caption{Spectral densities for a principal scalar with $\Delta = \frac{3}{2} + 0.7 i $ in $dS_4$ in complex $\omega$-plane. {The white spots mark the poles.}}
        \label{pic:rho}
\end{figure}
Therefore, for $t>0$ ($t<0$), closing the contour in the lower (upper) half plane, only the poles of $\tilde\rho^R(\omega)$ ($\tilde\rho^A(\omega)$) contribute, giving
\begin{align}
    \chi (t) =\theta(t) \int_{C_-} d
    \omega \, e^{-i\omega t}\tilde\rho^R(\omega) + \theta(-t) \int_{C_+} d
    \omega \, e^{-i\omega t} \tilde\rho^A(\omega) 
\end{align}
where $C_-$ and $C_+$ are contours enclosing all the poles in the lower-half and upper-half planes respectively. Summing over these poles with the explicit expressions \eqref{eq:resonancepoles}, we have
\begin{align}
     \chi (t) =\theta(t) \frac{e^{-\Delta t}+e^{-\bar\Delta t}}{(1-e^{-t})^d}+\theta(-t)\frac{e^{\Delta t}+e^{\bar\Delta t}}{(1-e^{t})^d} \;, 
\end{align}
reproducing \eqref{eq:HCcharacter}.


\section{Comments on the exceptional scalars}\label{sec:exceptional}

In this section, we study free scalars with masses
\begin{align}
    m_k^2=-k(k+d) \; , \qquad k=0, 1,2,\cdots \; ,
\end{align}
corresponding to scaling dimensions of non-positive integers
\begin{align}\label{eq:exceptionalD}
    \Delta = -k \; .
\end{align}
While $k=0$ corresponds to a massless scalar, the scalars for any $k\geq 1$ are all tachyonic. Classical theories of these scalars exhibit shift symmetries \cite{Bonifacio:2018zex}. From the $SO(1,d+1)$ representation theory point of view, \eqref{eq:exceptionalD} correspond to the so-called exceptional series I UIRs of  \cite{Sun:2021thf}, even though their realizations in terms of a local quantum field theory remain to be understood. Despite their tachyonic nature, they could be of physical interest because they generically appear in the multiparticle Hilbert spaces of massive fields \cite{Dobrev:1977qv,Repka}. In 2D ($d=1$), these UIRs coincide with the discrete series, whose Hilbert space realizations are explored in \cite{Anninos:2023lin}.

\subsection{Massless scalar}\label{sec:shift0}

We start with the case of $\Delta =0$ \cite{Allen:1985ux,Allen:1987tz}, for which the action is invariant under a constant shift
\begin{align}
    \phi\to \phi+ c\;. 
\end{align}
In sharp contrast with principal or complementary series, the resonance poles (supposedly) at
\begin{equation}\label{eq:masslessrefreq}
    i\omega^{\Delta=0}_{nl} = l+2n \qquad \text{or} \qquad i\omega^{\bar\Delta=d}_{nl}=d+l+2n \; ,
\end{equation}
largely coincide with the Matsubara frequencies \eqref{eq:worldtubeexpand}. Also, there is a mode with $l=n=0$ lying exactly at the origin on the complex $\omega$-plane. This drastically changes the analytic structure of the Jost function,
\begin{align}\label{eq:masslessWjost}
   W^{\rm Jost}_{l}(\omega) = \frac{ 2^{1+i\omega}\Gamma (1-i \omega )}{\Gamma \left(\frac{l-i \omega}{2}\right) \Gamma \left(\frac{d+l-i \omega}{2} \right)} \;,
\end{align}
and therefore of the correlators as well. On one hand, except at $l=n=0$, the first tower \eqref{eq:masslessrefreq} are not zeros of \eqref{eq:masslessWjost}, meaning they are not true resonance poles. On the other, the second tower are zeros of \eqref{eq:masslessWjost} in even $d$ but not odd $d$.\footnote{See also \cite{Brady:1999wd,Choudhury:2003wd,Natario:2004jd} for earlier work on quasinormal modes for massless scalars.} The occurrence of the would-be resonance poles (and the analogous phenomena for general $k\geq 0$) is reminiscent of the pole-skipping phenomena \cite{Grozdanov:2017ajz,Blake:2017ris,Blake:2018leo}.

One can also work out the worldline retarded function by an inverse transform,
\begin{align}
   G^R(t)= \frac{1}{{\rm Vol}(S^{d-1})}\int_{C^+} d\omega \, e^{-i\omega t} \mathcal{G}^R_{l=0}(\omega) \;,
\end{align}
where the contour $C^+$ lies above all the poles, including the one at $\omega =0$, as illustrated in Figure \ref{pic:contour}. Performing the integral then leads to 
\begin{align}\label{eq:k0GR}
    G^R (t) 
    =&\frac{\theta(t) }{{\rm Vol}(S^{d-1})}\left[ 1-\cos \left(\frac{\pi  d}{2}\right)  \frac{4 }{B\left(\frac{d}{2},\frac{d}{2} \right)}\frac{e^{-d t}}{d} \,_2F_1\left( d , \frac{d}{2} , 1 + \frac{d}{2}, e^{-2t} \right)\right]\;.
\end{align}
One can check that this is equivalent to taking the limit $\Delta\to 0$ in \eqref{eq:GRWL}. Similarly, one can work out the advanced function, where the integration contour $C^-$ must lie below the real line to avoid the pole at $\omega =0$. With these, we can write down the spectral function:
\begin{align}\label{eq:GCmassless}
    G^C (t) 
    =&-i\left(G^R (t) -G^A (t)  \right)\nn\\
    =&\frac{-i}{{\rm Vol}(S^{d-1})}\left[ 1- \cos \left(\frac{\pi  d}{2}\right)  \frac{4 }{B\left(\frac{d}{2},\frac{d}{2} \right)}\frac{e^{-d |t|}}{d}  \,_2F_1\left( d , \frac{d}{2} , 1 + \frac{d}{2}, e^{-2|t|} \right)\right]\;.
\end{align}
Notice the drastic difference between even and odd $d$: for the latter case (which includes the realistic $d=3$ case), the second decaying term completely vanishes, leaving only the constant term. 

Before we move on, we note that since $\mathcal{G}^A(\omega;x|y)=\mathcal{G}^R(-\omega;x|y)$, the (full) spectral function can be written as
\begin{align}
     G^C (t;x|y)  =& -i\left( G^R (t;x|y) - G^A (t;x|y) \right) 
     = \frac{1}{{\rm Vol}(S^{d-1})}\int_{C^+} d\omega \, \left( e^{-i\omega t} -e^{i\omega t} \right)\mathcal{G}^R(\omega;x|y) \; .
\end{align}
Now, we deform the contour $C^+ \to \mathbb{R}_\epsilon + C_\epsilon$ as follows:
\begin{figure}[H]
\centering
\begin{tikzpicture}[scale=1.5]

\draw[->] (-2,0) -- (2,0) node[right] {$\Re(\omega)$};
\draw[->] (0,-0.75) -- (0,1.25) node[above] {$\Im(\omega)$};

\draw[thick,red] (0.075,0.075) -- (-0.075,-0.075);
\draw[thick,red] (-0.075,0.075) -- (0.075,-0.075);

\draw[very thick,darkgreen,arrow] (-2,0.6) -- (2,0.6) node[xshift=-0.5cm,above]{$C^+$};

\draw[thick,blue,arrow] (-1.8,0) -- (-0.2,0) node[xshift=-0.25cm,below]{$-\epsilon$};
\draw[thick,blue,arrow] (-0.2,0) arc (180:0:0.2) node[xshift=0.25cm,yshift=0.25cm,above]{$C_\epsilon$};
\draw[thick,blue,arrow] (0.2,0) -- (1.8,0) node[xshift=-2.35cm,yshift=-0.05cm,below]{$\epsilon$};

\end{tikzpicture}
\caption{The deformed contour consists of a semi-circle $C_\epsilon$ of radius $\epsilon>0$ and $\mathbb{R}_\epsilon= \mathbb{R}\setminus (-\epsilon,\epsilon)$.}
        \label{pic:contour}
\end{figure}
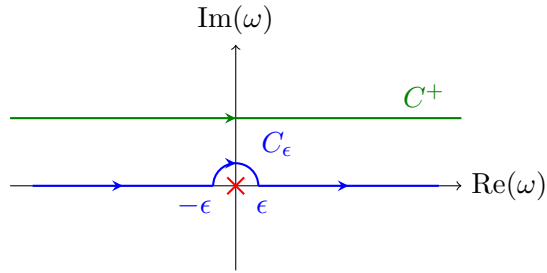
In the $\epsilon\to0$ limit, the contribution from the small semi-circle $C_\epsilon$ is given by $-\pi i$ times the residue at $\omega=0$, which is zero. To conclude, we have
\begin{align}\label{eq:GCmasslessintegral}
    G^C (t;x|y)  =  \frac{1}{{\rm Vol}(S^{d-1})} \dashint_{-\infty}^\infty d\omega \, e^{-i\omega t} \mathcal{G}^C(\omega;x|y) \;, 
\end{align}
where the integral is understood in the sense of the Cauchy principal value
\begin{align}\label{eq:Cauchy}
    \dashint_{-\infty}^\infty d\omega \equiv \lim_{\epsilon\to 0^+}\int_{\mathbb{R}\setminus (-\epsilon,\epsilon)} d\omega \;.
\end{align}

\subsubsection{Canonical quantization}

It is long known that the problem of quantizing a free massless scalar in de Sitter space is subtle. For example, the massive dS-invariant two-point function diverges in the massless limit, which is intimately tied with the fact that there is no dS-invariant Fock vacuum for a massless scalar \cite{Allen:1985ux}. However, there could exist vacua that preserve fewer symmetries, realized by quantizing the massless scalar in different patches \cite{Allen:1985ux,Allen:1987tz}. 

When we consider this problem in a static patch, we find that as long as we replace in \eqref{appeq:fieldop}
\begin{align}\label{eq:k0prescription}
    \int_{-\infty}^\infty d\omega\to \dashint_{-\infty}^\infty d\omega 
\end{align}
as defined in \eqref{eq:Cauchy}, the canonical quantization procedure in appendix \ref{sec:canquan} applies. In particular, we have a consistent {\it quantum} theory that reproduces correlators such as \eqref{eq:k0GR} and \eqref{eq:GCmassless}. Subsequently, we can also define the temperature-sensitive functions. As a demonstration, we compute the symmetric Wightman function at the dS temperature $\beta=\beta_{\rm dS}=2\pi$:
\begin{align}\label{eq:symGCtk0}
    G_{\beta = 2\pi}^{\rm Sym} (t) = \frac{1}{{\rm Vol}(S^{d-1})}\dashint_{-\infty}^\infty d\omega \, e^{-i\omega t}\mathcal{G}^{\rm Sym}_{l=0} (2\pi,\omega)\; ,
\end{align}
where 
\begin{align}\label{eq:k0Gsym}
    \mathcal{G}^{\rm Sym}_{l=0} (2\pi,\omega) =\frac{1}{8 \pi^2} \frac{1}{\Gamma\left(\frac{d}{2}\right)^2}
    \cosh \pi \omega \left| \Gamma\left( \frac{-i\omega}{2}\right) \Gamma\left( \frac{d-i\omega}{2}\right)  \right|^2 \;. 
\end{align}
We will evaluate \eqref{eq:symGCtk0} explicitly and show that it can be obtained as an analytic continuation of an appropriate $S^{d+1}$ correlator.

Clearly \eqref{eq:k0Gsym} has poles at the resonances \eqref{eq:masslessrefreq} with $l=0$,
\begin{align}\label{eq:k0l0resonance}
    i\omega^{\Delta=0}_{n}=2n\; , \qquad i\omega^{\bar\Delta=d}_{n}=d+2n \; ,
\end{align}
and the anti-resonances given by $-i\omega^{\Delta=0}_{n}$ and $-i\omega^{\bar\Delta=d}_{n}$. In any dimension, the pole at $\omega=\omega^{\Delta=0}_{n=0}=0$ is a {\it double} pole, where the integrand \eqref{eq:symGCtk0} has residue 
\begin{align}
    -\pi i\operatorname*{Res}\limits_{\omega= 0}  e^{-i\omega t}\mathcal{G}^{\rm Sym}_{l=0} (2\pi,\omega) = -\frac{t}{2\pi} \; . 
\end{align}
To proceed, we separate the cases of odd and even $d$.

\paragraph{When $d$ is odd}

Except for the double pole at $\omega =0$, both towers \eqref{eq:k0l0resonance} are simple poles for the integrand \eqref{eq:symGCtk0}, at which the residues are
\begin{align}
    -2\pi i \operatorname*{Res}\limits_{\omega= -i 2n}  e^{-i\omega t}\mathcal{G}^{\rm Sym}_{l=0} (2\pi,\omega) = \frac{(-1)^n  }{2 \pi  \Gamma \left(\frac{d}{2}\right)^2}  \Gamma \left(\frac{d}{2}-n\right) \Gamma\left(\frac{d}{2}+n\right) \frac{e^{-2 n t}}{n} \; , \qquad n\geq 1 \; ,
\end{align}
and 
\begin{align}
    -2\pi i\operatorname*{Res}\limits_{\omega= -i(d+2n)}  e^{-i\omega t}\mathcal{G}^{\rm Sym}_{l=0} (2\pi,\omega) =
    \frac{(-1)^{\frac{d-1}{2}}}{ \Gamma
   \left(\frac{d}{2}\right)^2} \frac{ \Gamma (d+n)}{ d+2 n  }\frac{e^{- (d+2 n)t}}{n!} \; , \qquad n\geq 0 \; ,
\end{align}
respectively. Therefore, for $t>0$ ($t<0$), closing the contour in the lower (upper) half plane, we sum over the residues and obtain
\begin{align}\label{eq:k0symodd}
    G_{\beta = 2\pi}^{\rm Sym} (t) =& \frac{1}{{\rm Vol}(S^{d-1})}\bigg[ -\frac{|t|}{2 \pi } -\frac{d }{2 \pi  (d-2)} e^{-2 |t|}  \, _3F_2\left(1,1,\frac{d}{2}+1;2,2-\frac{d}{2};e^{-2 |t|}\right)\nn\\
    &+\frac{(-)^{\frac{d-1}{2}} }{B\left(\frac{d}{2},\frac{d}{2} \right)}\frac{e^{-d |t| }}{d}\, _2F_1\left( d, \frac{d}{2}, 1 + \frac{d}{2} , e^{-2|t|} \right) \bigg] \;.
\end{align}
We note that the generalized hypergeometric function in the first line can be written as \cite{Hypergeometric2F1_2024}
\begin{align}
    e^{-2 |t|}  \, _3F_2\left(1,1,\frac{d}{2}+1;2,2-\frac{d}{2};e^{-2 |t|}\right)  = - \frac{d-2}{d}\left[ \frac{\partial}{\partial \Delta} {}_2 F_1\left( \Delta , \frac{d}{2}; 1- \frac{d}{2}; e^{-2t} \right)\Big|_{\Delta \rightarrow 0} \right] \,.
\end{align}
Therefore, 
\begin{align}
    G_{\beta = 2\pi}^{\rm Sym} (t)
    =& \frac{1 }{{\rm Vol}(S^{d-1})} \bigg[-\frac{|t|}{2\pi}+\frac{1}{2\pi}\partial_\Delta \, _2 F_1\left( \Delta , \frac{d}{2}; 1- \frac{d}{2}; e^{-2|t|} \right)\Big|_{\Delta =0} \nn\\
    &+\frac{(-)^{\frac{d-1}{2}} }{B\left(\frac{d}{2},\frac{d}{2} \right)}\frac{e^{-d |t| }}{d}\, _2F_1\left( d, \frac{d}{2}, 1 + \frac{d}{2} , e^{-2|t|} \right) \bigg]\;. 
\end{align}

\paragraph{When $d$ is even}

The integrand \eqref{eq:symGCtk0} has simple poles at the first tower of \eqref{eq:k0l0resonance} for $1\leq n\leq \frac{d}{2}-1$, at which it has residues 
\begin{align}
     -2\pi i \operatorname*{Res}\limits_{\omega= -i 2n}  e^{-i\omega t}\mathcal{G}^{\rm Sym}_{l=0} (2\pi,\omega) = \frac{(-1)^n  }{2 \pi  \Gamma \left(\frac{d}{2}\right)^2}  \Gamma \left(\frac{d}{2}-n\right) \Gamma\left(\frac{d}{2}+n\right) \frac{e^{-2 n t}}{n}  \; , \quad 1\leq n\leq \frac{d}{2}-1\; . 
\end{align}
For all $n\geq \frac{d}{2}$, the integrand \eqref{eq:symGCtk0} has double poles instead, at which it has residues 
\begin{align}\label{eq:k0evendresidue}
    &-2\pi i\operatorname*{Res}\limits_{\omega= -i 2n}  e^{-i\omega t}\mathcal{G}^{\rm Sym}_{l=0} (2\pi,\omega) \nn\\
    &= \frac{(-)^{d/2}}{2 \pi  \Gamma \left(\frac{d}{2}\right)^2}\frac{\Gamma \left(\frac{d}{2}+n\right) }{ n}\left( \frac{1}{n}+2 t+\psi\left(n-\frac{d}{2}+1\right)-
   \psi\left(n+\frac{d}{2}\right)\right)\frac{e^{-2nt}}{\Gamma
   \left(n-\frac{d}{2}+1\right)} \; , \quad n\geq \frac{d}{2}\; .
\end{align}
In this case, the second tower \eqref{eq:k0l0resonance} completely overlaps with the $n\geq \frac{d}{2}$ part of the first tower and hence does not lead to separate contributions.

Therefore, for $t>0$ ($t<0$), closing the contour in the lower (upper) half plane, we have 
\begin{align}
    G_{\beta = 2\pi}^{\rm Sym} (t)
    =& \frac{-1}{{\rm Vol}(S^{d-1})} \left[ \frac{|t|}{2 \pi } +2\pi i\left( \sum_{n=0}^{\frac{d}{2}-1} + \sum_{n=\frac{d}{2}}^\infty \right)\operatorname*{Res}\limits_{\omega= -i 2n}  e^{-i\omega |t|}\mathcal{G}^{\rm Sym}_{l=0} (2\pi,\omega) \right] \; .
\end{align}
Shifting the second sum by $n\to n+\frac{d}{2}$, we have
\begin{align}\label{eq:k0symeven}
    &G_{\beta = 2\pi}^{\rm Sym} (t) \nn\\
    =& \frac{1}{{\rm Vol}(S^{d-1})}  \bigg[-\frac{|t|}{2 \pi } +\frac{1}{B\left(\frac{d}{2},\frac{d}{2} \right)} \sum_{n=0}^{\frac{d}{2}-1}\frac{ (-1)^n  }{2 \pi  \Gamma \left(d\right)}  \Gamma \left(\frac{d}{2}-n\right) \Gamma\left(\frac{d}{2}+n\right) \frac{e^{-2 n |t|}}{n}\nn\\
     &+  \frac{(-)^{d/2}}{B\left(\frac{d}{2},\frac{d}{2} \right)} \frac{e^{-d|t|}}{\pi }\sum_{n=0}^{\infty}\frac{1}{ \Gamma \left(d\right)} \frac{\Gamma \left(d+n\right)}{d+2n}
   \left( \frac{2}{d+2n}+2|t|+\psi\left(n+1\right)-
   \psi\left(n+d\right)\right)\frac{e^{-2n|t|}}{n!} \bigg] \;.
\end{align}
In the second equality, we have shifted the sum by $n\to n+\frac{d}{2}$.

\subsubsection{Euclidean correlator}

A natural question is whether the real-time thermal correlators computed above correspond to some analytic continuation from a $S^{d+1}$ correlator. This is not as trivial as one might have thought, because in contrast to the massive case, a massless scalar does not have a standard Green function satisfying \eqref{eq:sphereGreeneom}; the Laplacian has a normalizable zero mode and is thus non-invertible. It is also reflected in the fact that the massive expression \eqref{eq:EuclideanGreen} diverges as $\Delta \to 0$. An object we can consider is a {\it modified} Green function that satisfies 
 \begin{equation}\label{eq:excepSphereGreen}
     -\nabla_\mathbf{x}^2 \hat G_E(P(\mathbf{x},\mathbf{y})) =  \frac{1}{\sqrt{g}}\delta^{(d+2)}(\mathbf{x},\mathbf{y}) -\frac{1}{\text{Vol} (S^{d+1})} \; .
\end{equation}
Because of the constant term on the right hand side, it is an {\it inhomogeneous} equation even at non-coincident points. This constant term represents a non-local constraint of the problem, analogous to the Gauss law constraint that forbids putting a single charge on a sphere. A particular solution to \eqref{eq:excepSphereGreen} is given by \cite{Epstein:2014jaa}
\begin{align}\label{eq:modifiedGE}
    \hat{\mathbf{G}}(\mathbf{x},\mathbf{y})  \equiv \frac{d}{d\Delta}\left[\frac{G^\Delta_E\left(P(\mathbf{x},\mathbf{y})\right)}{\Gamma(\Delta)} \right]\Big|_{\Delta=0} \; ,
\end{align}
with $G^\Delta_E\left(P(\mathbf{x},\mathbf{y})\right)$ being the massive Green function \eqref{eq:EuclideanGreen}. One can see that \eqref{eq:modifiedGE} is a solution to \eqref{eq:excepSphereGreen} by dividing \eqref{eq:sphereGreeneom} by $\Gamma(\Delta)$ and evaluating its $\Delta$-derivative at $\Delta =0$. Notice that \eqref{eq:modifiedGE} is not a unique solution to \eqref{eq:excepSphereGreen} since we can always add a constant $C$:
\begin{align}\label{eq:k0Greengen}
    \hat G_E\left(P(\mathbf{x},\mathbf{y})\right) = \hat{\mathbf{G}}\left(P(\mathbf{x},\mathbf{y})\right) + C \; . 
\end{align}
In the following we will demonstrate that upon choosing an appropriate constant $C$, the unique $\hat G_E(P(\mathbf{x},\mathbf{y}))$ would analytically continue to the real-time correlators as in the massive case.

The modified function \eqref{eq:modifiedGE} is convenient to work with since we can simply take limits in the previous massive results. We can apply the prescription \eqref{eq:modifiedGE} directly to the analytically continued expression \eqref{eq:continuedGE}, whose second term is finite for odd $d$ and diverges for even $d$ as $\Delta\to 0$.

\paragraph{When $d$ is odd}

Applying \eqref{eq:modifiedGE} to \eqref{eq:continuedGE} yields
\begin{align}
	\hat{\mathbf{G}}(P\mp i \epsilon) 
 =& \frac{1 }{2{\rm Vol}(S^{d+1})\Gamma\left( \frac{d}{2}+1\right)} \partial_\Delta \left[ e^{\mp i \pi \Delta} \Gamma\left(\frac{d}{2}-\Delta\right)e^{-\Delta |t| }\, _2F_1\left( \Delta, \frac{d}{2}, 1 - \frac{d}{2} + \Delta, e^{-2|t|} \right)\right]_{\Delta=0} \nn\\
 &-\frac{ \Gamma(d)\Gamma(-\frac{d}{2}) }{2{\rm Vol}(S^{d+1})\Gamma\left( \frac{d}{2}+1\right)}e^{-d |t| }\, _2F_1\left( d, \frac{d}{2}, 1 + \frac{d}{2} , e^{-2|t|} \right) \; .
\end{align}
Using the series representation for the hypergeometric function
\begin{align}\label{eq:2f1series}
    \, _2F_1\left(a,b, c, z \right) = \sum_{n=0}^\infty \frac{\left(a\right)_n \left(b\right)_n}{\left(c\right)_n}\frac{z^n}{n!} \; ,
\end{align}
the derivative on the first line can be computed to be 
\begin{align}
    &\partial_\Delta \left[ e^{\mp i \pi \Delta} e^{-\Delta |t| } \Gamma\left(\frac{d}{2}-\Delta\right)\, _2F_1\left( \Delta, \frac{d}{2}, 1 - \frac{d}{2} + \Delta, e^{-2|t|} \right)\right]_{\Delta=0}\nn\\
    &=\Gamma\left(\frac{d}{2}\right)\left[ \mp i \pi  - |t|  +\partial_\Delta \, _2 F_1\left( \Delta , \frac{d}{2}; 1- \frac{d}{2}; e^{-2|t|} \right)\Big|_{\Delta =0}- \psi\left(\frac{d}{2}\right)\right] \;. 
\end{align}
Here $\psi(z)=\partial_z \log \Gamma(z)$ is the digamma function. Choosing
\begin{align}
    C =\frac{1 }{2\pi {\rm Vol}(S^{d-1})}\psi\left(\frac{d}{2}\right)
\end{align}
in \eqref{eq:k0Greengen}, i.e. 
\begin{align}
     \hat G_E(P(\mathbf{x},\mathbf{y})) =& \hat{\mathbf{G}}\left(P(\mathbf{x},\mathbf{y})\right) + \frac{1 }{2\pi {\rm Vol}(S^{d-1})}\psi\left(\frac{d}{2}\right) \qquad (\text{odd } d)\; ,
\end{align}
we have the {\it unique} sphere correlator that analytically continues to the Lorentzian signature as in the massive case \eqref{eq:massanacno}, i.e.
\begin{align}
    G^C (t) &= \hat G_E(P- i \epsilon) - \hat G_E(P+i \epsilon) \; , \qquad 
    G_{\beta = 2\pi}^{\rm Sym} (t) = \frac{\hat G_E(P- i \epsilon) + \hat G_E(P+i \epsilon)}{2}\;,
\end{align}
where $G^C (t)$ and $G_{\beta = 2\pi}^{\rm Sym} (t)$ are defined in \eqref{eq:GCmassless} and \eqref{eq:k0symodd} respectively.

\paragraph{When $d$ is even}

Applying \eqref{eq:modifiedGE} to \eqref{eq:continuedGE} yields 
\begin{align}\label{eq:k0Euevend}
	&\hat{\mathbf{G}}(P\mp i \epsilon) \nn\\
 =& \frac{1}{2{\rm Vol}(S^{d+1})\Gamma\left( \frac{d}{2}+1\right)}\bigg( \partial_\Delta \left[ e^{\mp i \pi \Delta} \Gamma\left(\frac{d}{2}-\Delta\right)e^{-\Delta |t| }\, _2F_1\left( \Delta, \frac{d}{2}, 1 - \frac{d}{2} + \Delta, e^{-2|t|} \right)\right]_{\Delta=0} \nn\\
 &+\partial_\Delta \left[ e^{\mp i \pi \bar\Delta} \frac{\Gamma(\bar\Delta)\Gamma\left(\frac{d}{2}-\bar\Delta\right)}{\Gamma(\Delta)}e^{-\bar\Delta |t| }\, _2F_1\left( \bar\Delta, \frac{d}{2}, 1 - \frac{d}{2} + \bar\Delta, e^{-2|t|} \right)\right]_{\Delta=0}\bigg) \; .
\end{align}
The computation in this case is more subtle. For example, when using \eqref{eq:2f1series} to compute the first line, we note that in the $\Delta\to 0$ limit, the term $n=0$ survives, and those with $1\leq n\leq \frac{d}{2}-1$ go to zero, while all terms with $n\geq \frac{d}{2}$ acquire a non-trivial limit:
\begin{align}
    \lim_{\Delta \to 0}\, _2F_1\left( \Delta, \frac{d}{2}, 1 - \frac{d}{2} + \Delta, e^{-2|t|} \right) 
    =&1+ \frac{2(-)^{\frac{d}{2}+1}}{B \left(\frac{d}{2},\frac{d}{2}\right)}e^{-d |t|}\sum_{n=0}^\infty \frac{1}{\Gamma(d)}\frac{\Gamma\left(d+n\right)  }{d+2n}\frac{e^{-2n|t|}}{n!} \; . 
\end{align}
After some computation with all limits carefully taken care of, we eventually obtain
\begin{align}
    &\hat{\mathbf{G}}(P\mp i \epsilon) \nn\\
    =&\frac{1 }{ {\rm Vol}(S^{d-1})}\bigg[\mp \frac{i}{2}-\frac{|t|}{2\pi} +\frac{1}{B\left( \frac{d}{2},\frac{d}{2}\right)}\sum_{n=0}^{\frac{d}{2}-1} \frac{(-)^n}{2\pi\Gamma(d)} \Gamma\left(\frac{d}{2}+n\right)  \Gamma\left(\frac{d}{2} - n \right) \frac{e^{-2n|t|}}{n}\nn\\
    &+ \frac{(-)^{\frac{d}{2}}}{B\left( \frac{d}{2},\frac{d}{2}\right)}\frac{e^{-d|t|}}{\pi}\sum_{n=0}^\infty \frac{ 1}{\Gamma\left(d\right)}\frac{\Gamma\left(d+n\right)}{d+2n} \left(\mp i 2\pi + \frac{2}{d+2n}+2|t|+\psi\left(n+1\right)-
   \psi\left(n+d\right)\right)\frac{e^{-2n|t|}}{n!}  \bigg] \;.
\end{align}
Choosing $C =0$ in \eqref{eq:k0Greengen}, i.e. 
\begin{align}
     \hat G_E(P(\mathbf{x},\mathbf{y})) = \hat{\mathbf{G}}\left(P(\mathbf{x},\mathbf{y})\right)   \qquad (\text{even } d)\; ,
\end{align}
we have the {\it unique} sphere correlator that analytically continues to the Lorentzian signature as in the massive case \eqref{eq:massanacno}, i.e.
\begin{align}
    G^C (t) &= \hat G_E(P- i \epsilon) - \hat G_E(P+i \epsilon) \; , \qquad 
    G_{\beta = 2\pi}^{\rm Sym} (t) = \frac{\hat G_E(P- i \epsilon) + \hat G_E(P+i \epsilon)}{2}\;,
\end{align}
where $G^C (t)$ and $G_{\beta = 2\pi}^{\rm Sym} (t)$ are defined in \eqref{eq:GCmassless} and \eqref{eq:k0symeven} respectively. 

To summarize, we have demonstrated that the prescription \eqref{eq:k0prescription} leads to a consistent quantization for a massless scalar in the static patch, and performed a non-trivial check that the Lorentzian correlators can all be analytically continued from an appropriate $S^{d+1}$ correlator. However, this is unlikely the full story. Specifically, the prescription \eqref{eq:k0prescription} explicitly excludes the constant/soft mode $\omega=0$, which could encode interesting physics.\footnote{In the context of Minkowski space, the constant/soft modes play a crucial role in geometric soft theorems \cite{Cheung:2021yog,Kapec:2022axw}.} For example, the constant mode seems closely related to the fact that extra data is required to fully determine the massless scalar 1-loop $S^{d+1}$ partition functions \cite{Law:2020cpj}, such as the target space geometry if the parent interacting theory is a non-linear sigma model.

\subsection{General shift-symmetric scalars}\label{sec:shiftk}

For any $\Delta =- k , k\geq 1$, the scalars are tachyonic. See for example \cite{Bros:2010wa,Epstein:2014jaa} for the study of such tachyonic scalars. Their free actions are invariant under the shift symmetries
\begin{align}\label{eq:kshiftsym}
    \phi\to \phi+ c f_k \;,
\end{align}
where $f_k$ is a homogeneous polynomial in ambient space coordinates of degree $k$ \cite{Bonifacio:2018zex}. The $k=1$ and $k=2$ cases are the dS analogs for the Galileon and special Galileon theories in flat space \cite{Bonifacio:2021mrf}, respectively. The former case can also be obtained by truncating linearized gravity to its conformal sector, and thus has been studied as a toy model for the latter \cite{Folacci:1992}.

\paragraph{Causal correlators}

Similar to the massless case, the resonance poles (supposedly) at
\begin{equation}\label{eq:shiftkrefreq}
    i\omega^{\Delta=-k}_{nl} = -k+l+2n \qquad \text{or} \qquad i\omega^{\bar\Delta=d+k}_{nl}=d+k+l+2n 
\end{equation}
largely coincide with the Matsubara frequencies \eqref{eq:worldtubeexpand}. In addition to the modes with $\omega=0$ (when $l+2n=k$) lying exactly at the origin, there are a finite number of poles lying {\it above} the real line, namely those in the first tower with
\begin{align}\label{eq:shiftedpoles}
    l+2n<k \;.
\end{align}
The presence of these modes indicates a kind of superradiant instability, typical in tachyonic theories. Similar to the massless case, except when $l+2n\leq k$, the first tower \eqref{eq:masslessrefreq} are not zeros of the Jost function,
\begin{align}\label{eq:shiftkWjost}
   W^{\rm Jost}_{l}(\omega) = \frac{ 2^{1+i\omega}\Gamma (1-i \omega )}{\Gamma \left(\frac{-k+l-i \omega}{2}\right) \Gamma \left(\frac{d+k+l-i \omega}{2} \right)} \; , 
\end{align}
while the second tower is zeros of \eqref{eq:shiftkWjost} in even $d$ but not odd $d$. 

In computing the worldline retarded function,
\begin{align}\label{eq:shiftkGR}
   G^R(t)= \frac{1}{{\rm Vol}(S^{d-1})}\int_C d\omega \, e^{-i\omega t} \mathcal{G}^R_{l=0}(\omega) \; , 
\end{align}
we must choose the contour lie above the highest pole $\omega =ik$. The result is equivalent to taking the limit $\Delta\to-k$ in \eqref{eq:GRWL}:
\begin{align}\label{eq:GRWLshiftk}
    G^R (t) =&\frac{\theta(t)}{{\rm Vol}(S^{d-1})} \bigg[\frac{ 1 }{B\left(\frac{d}{2},k \right)} \frac{e^{k t}}{k}  \,_2F_1\left( -k , \frac{d}{2} , 1 - \frac{d}{2} -k, e^{-2t} \right) \nn\\
    &-\cos \left(\frac{\pi  d}{2}\right)  \frac{4 }{B\left(\frac{d}{2},\frac{d}{2} +k \right)}\frac{e^{-(d+k) t}}{d+2k} \,_2F_1\left( d +k, \frac{d}{2} , 1 + \frac{d}{2}+k, e^{-2t} \right) \bigg] \; .
\end{align}
The advanced function can be similarly obtained. Taking the difference, we have the spectral function: 
\begin{align}\label{eq:GCWLshiftk}
    G^C (t) =&\frac{-i}{{\rm Vol}(S^{d-1})} \bigg[\frac{ 1 }{B\left(\frac{d}{2},k \right)} \frac{e^{k |t|}}{k}  \,_2F_1\left( -k , \frac{d}{2} , 1 - \frac{d}{2} -k, e^{-2|t|} \right) \nn\\
    &-\cos \left(\frac{\pi  d}{2}\right)  \frac{4 }{B\left(\frac{d}{2},\frac{d}{2} +k \right)}\frac{e^{-(d+k) |t|}}{d+2k} \,_2F_1\left( d +k, \frac{d}{2} , 1 + \frac{d}{2}+k, e^{-2|t|} \right) \bigg] \;.
\end{align}

\paragraph{Comments on quantization}

Notice that \eqref{eq:GRWLshiftk} or \eqref{eq:GCWLshiftk} implies that the {\it full} correlators {\it grow} indefinitely for two points at large timelike separations. It is not possible to reproduce these by a quantum free real scalar theory with usual properties, namely that we have a self-adjoint field operator $\hat \phi(t)$ (suppressing the spatial dependence) that evolves in time unitarily as $\hat \phi(t)=e^{i \hat Ht }\hat \phi(0)e^{-i \hat Ht }$, generated by a self-adjoint Hamiltonian $\hat H$ with respect to which we have a normalizable vacuum state $\ket{0}$. Otherwise, consider the smeared field operator 
\begin{align}\label{eq:smearing}
    \hat \phi_f (T)= \int dt \,dx\, d\Omega\,  \hat \phi (t+T,x,\Omega) \, f(t,x,\Omega)
\end{align}
for some smooth function $f(t,x,\Omega)$ on the static patch, e.g. a Gaussian centered around $t=0$ and $x=0$. Then, the Wightman function for this smeared operator at general $T>0$ must satisfy\footnote{The smearing \eqref{eq:smearing} is to guarantee that $\phi_f(0)\ket{0}$ has a finite norm.}
\begin{align}\label{eq:shiftargument}
    \bra{0} \phi_f (T)\phi_f(0)\ket{0}
    \leq &\sqrt{\bra{0} \phi_f(t) \phi_f(t)\ket{0}  } \sqrt{\bra{0} \phi_f(0) \phi_f(0)\ket{0}  }= \bra{0} \phi_f(0) \phi_f(0)\ket{0} 
    \; , 
\end{align}
i.e. it is bounded from above by its value at $T=0$. Since all other two-point functions are linear combinations of the Wightman function, their time evolutions are bounded as well. This clearly contradicts the exponentially growing behaviors of \eqref{eq:GRWLshiftk} or \eqref{eq:GCWLshiftk}, since the smeared functions behave like the un-smeared ones for $T$ much greater than the width of the function $f$.

This is consistent with the existing discussions in the literature: to have a self-consistent quantization, the shift symmetry \eqref{eq:kshiftsym} needs to be gauged in some way -- coupling these scalars with gauge fields \cite{Anninos:2023lin}, for example, so that the modes responsible for the growing behaviors are pure-gauge.

\subsection{Harish-Chandra characters for the exceptional scalars}\label{ref:excepchar}

In this section, we analyze the structures of the Harish-Chandra characters for the exceptional scalars along the lines of section \ref{sec:charpole}. To that end, we stress that the analysis for the Klein-Gordon equation for the exceptional scalars are no different from the generic massive case, which implies that the discussion in section \ref{sec:Smatrix} remains valid. In particular, the S-matrices for the exceptional case are given by taking $\Delta\to-k$ in \eqref{eq:dSSmat}:   
\begin{gather}\label{eq:dSSmatk}
    \mathcal{S}_l (\omega)  = \mathcal{S}_l^\text{dS} (\omega) \mathcal{S}^\text{Rin} \left(\omega\right)\;,\nn\\
    \mathcal{S}_l^\text{dS} (\omega)\equiv \frac{\Gamma\left(\frac{-k+l- i  \omega}{2}\right) \Gamma\left(\frac{d+k+l- i  \omega}{2}\right)}{\Gamma\left(\frac{-k+l+i  \omega}{2}\right) \Gamma\left(\frac{d+k+l+ i \omega}{2}\right)}  \; , \qquad 
    \mathcal{S}^\text{Rin} \left(\omega\right) =  2^{-2i\omega}  \frac{\Gamma(i\omega)}{\Gamma(-i\omega)} \; .
\end{gather}
As discussed in section \ref{sec:shift0} and \ref{sec:shiftk}, a subset of \eqref{eq:masslessrefreq} or \eqref{eq:shiftkrefreq} are not true resonance poles for the Klein-Gordon equation, since they are not zeros of the Jost function. However, they are poles of $\mathcal{S}_l^\text{dS} (\omega)$ (and therefore of $\tilde\rho^R(\omega)$). Because of this, we will abuse the language in this section and call them the resonance poles. Similarly, we will call the zeros of $\mathcal{S}_l^\text{dS} (\omega)$ (poles of $\tilde\rho^A(\omega)$) the anti-resonance poles.

\paragraph{Massless scalar}

\begin{figure}[H]
    \centering
   \begin{subfigure}{0.3\textwidth}
            \centering
            \includegraphics[height=4.0cm]{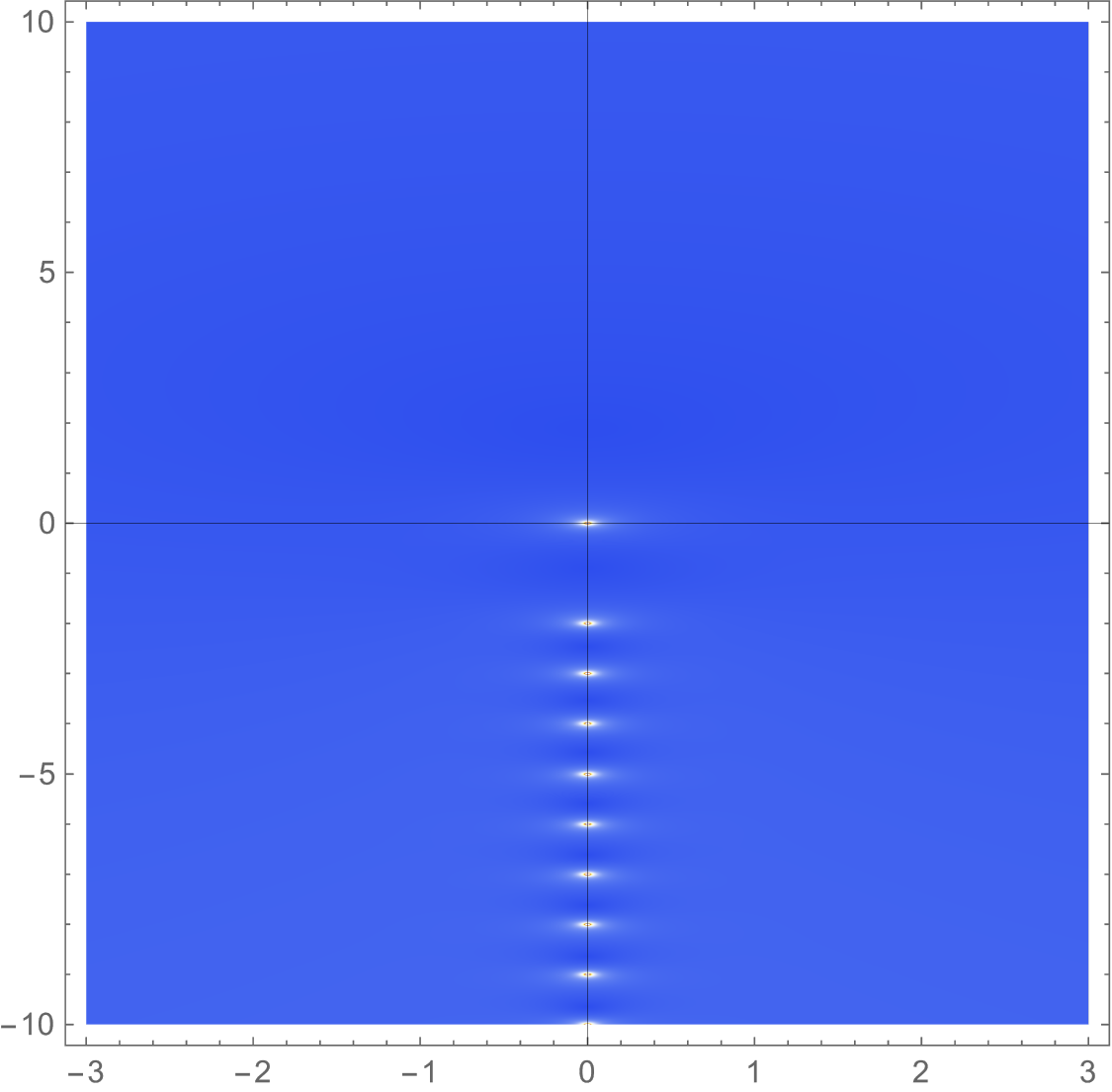}
            \caption[]%
            {{$\left|\Delta \rho^R_{l=0}(\omega)\right|$}}    
        \end{subfigure}
        \hspace{0.25cm}
        \begin{subfigure}{0.3\textwidth}  
            \centering 
            \includegraphics[height=4.0cm]{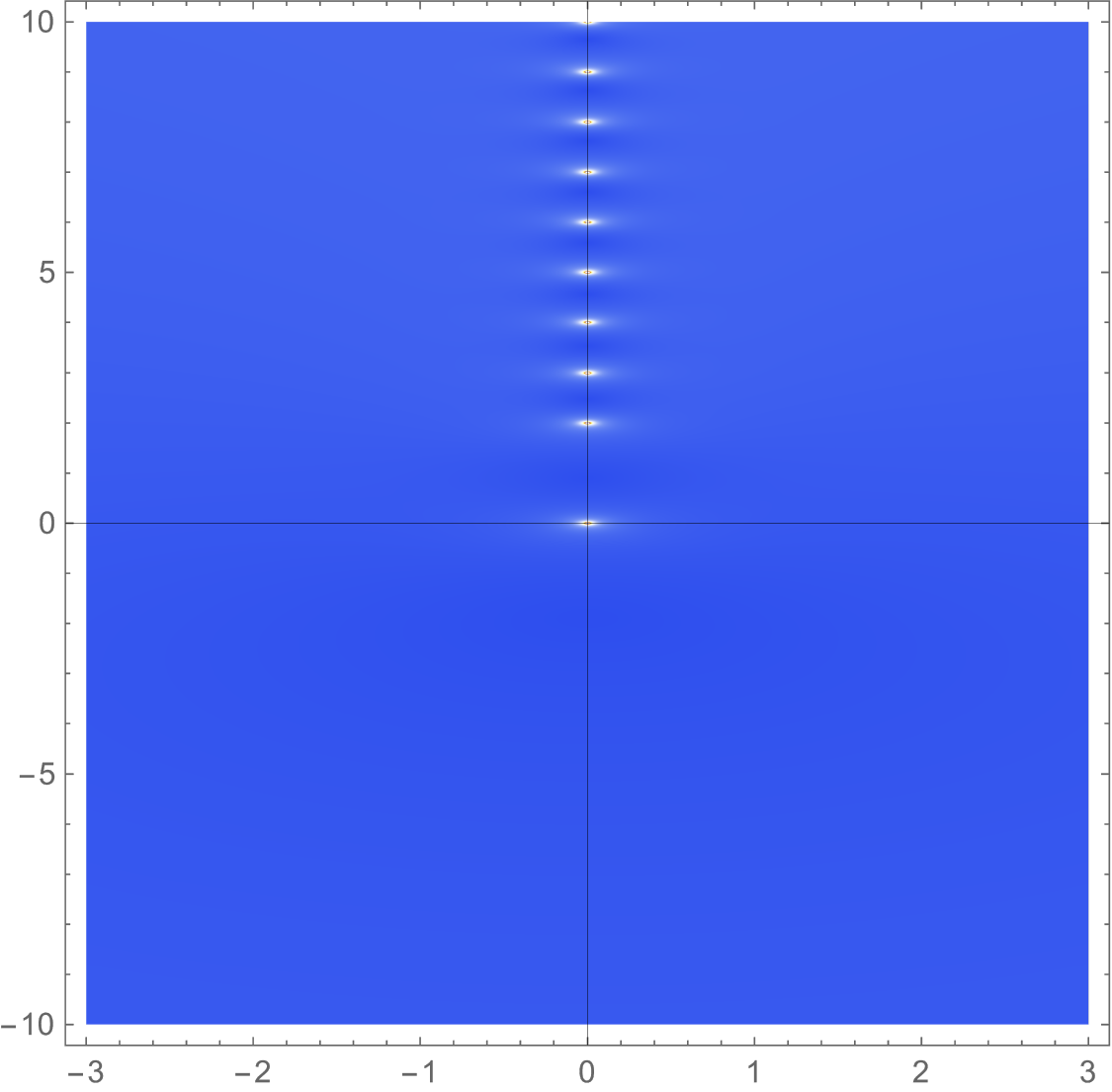}
            \caption[]%
            {{$\left|\Delta \rho^A_{l=0}(\omega)\right|$}}    
        \end{subfigure}
        \hspace{0.25cm}
        \begin{subfigure}{0.3\textwidth}  
            \centering 
            \includegraphics[height=4.0cm]{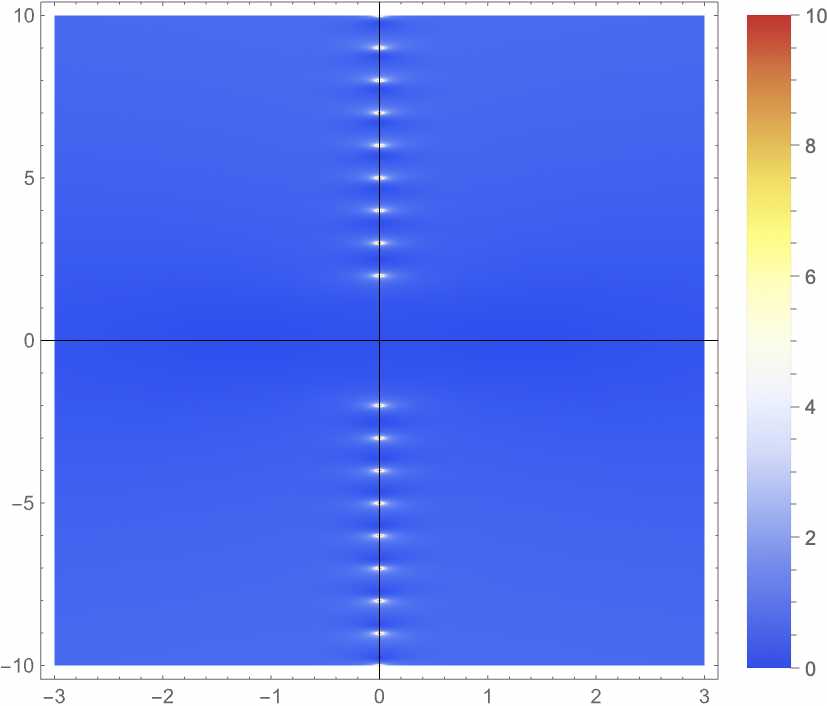}
            \caption[]%
            {{\small \centering  $\left|\Delta \rho_{l=0}(\omega)\right|$}}    
        \end{subfigure}
        \caption{Spectral densities for a massless scalar in $dS_4$ in complex $\omega$-plane. { The white spots mark the poles.}}
        \label{pic:rhok0}
\end{figure}

As we saw in section \ref{sec:shift0}, in the massless case the (anti-)resonance poles do not all lie in the lower (upper) half plane. In particular, there is a pole for $\tilde\rho^R(\omega)$ and $\tilde\rho^A(\omega)$ with exactly zero frequency $\omega=0$ ($n=l=0$ in \eqref{eq:masslessrefreq}). See figure \ref{pic:rhok0}. In the total relative spectral density \eqref{eq:dossplit}, the contributions from these two poles cancel each other. 

Therefore, closing the contour in the lower (upper) half plane for $t>0$ ($t<0$), the inverse transform \eqref{eq:chiinverse} receives contributions from all (anti-)resonance frequencies {\it except} $\omega=0$, yielding 
\begin{align}
     \chi (t) = \frac{1+e^{-d t}}{|1-e^{-t}|^d}-1 \; . 
\end{align}
The subtraction by 1 accounts for the fact that the (anti-)resonance pole $\omega=0$ does not contribute.

\paragraph{General exceptional scalars}

\begin{figure}[H]
    \centering
   \begin{subfigure}{0.3\textwidth}
            \centering
            \includegraphics[height=4.0cm]{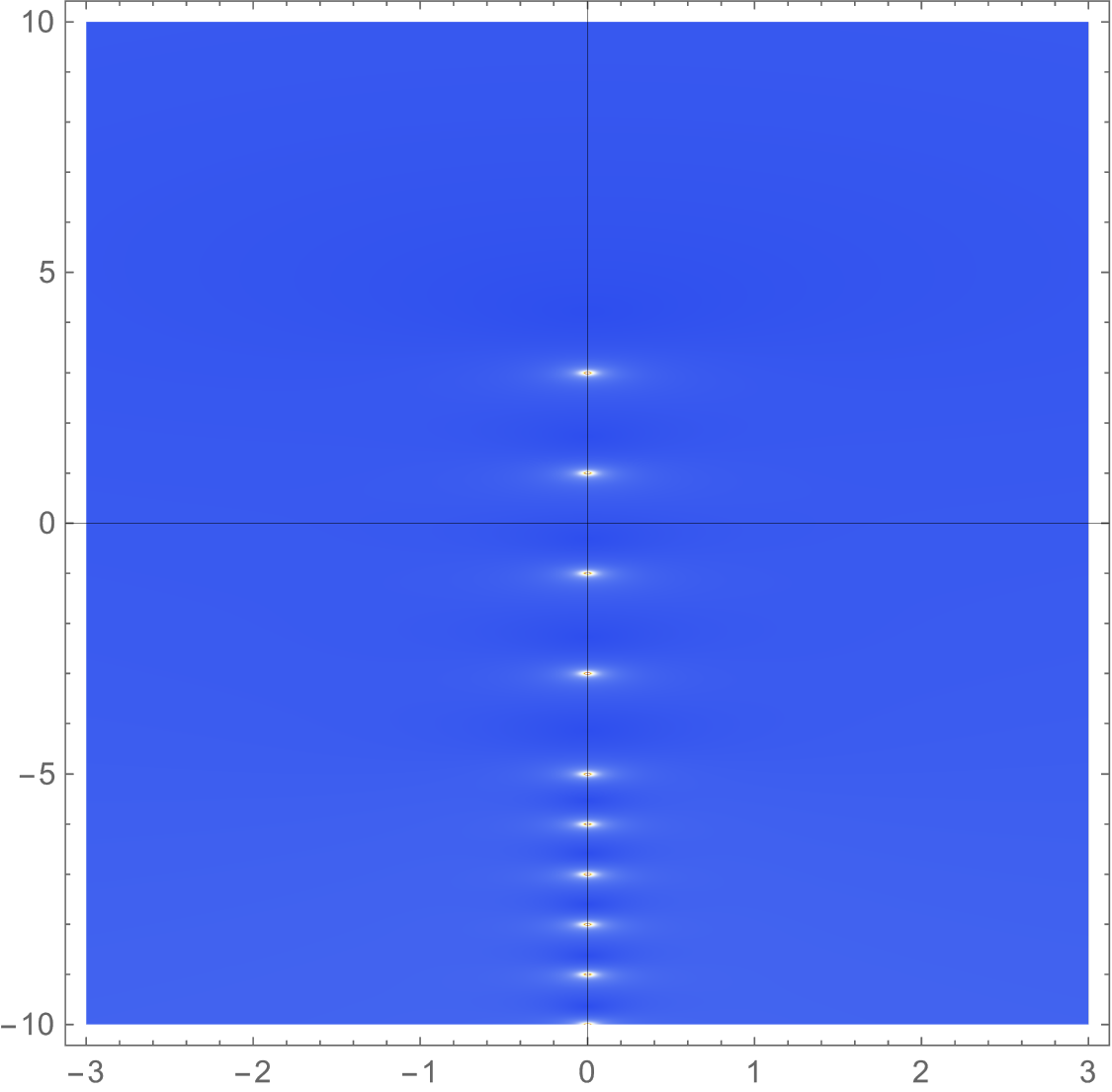}
            \caption[]%
            {{$\left|\Delta \rho^R_{l=0}(\omega)\right|$}}    
        \end{subfigure}
        \hspace{0.25cm}
        \begin{subfigure}{0.3\textwidth}  
            \centering 
            \includegraphics[height=4.0cm]{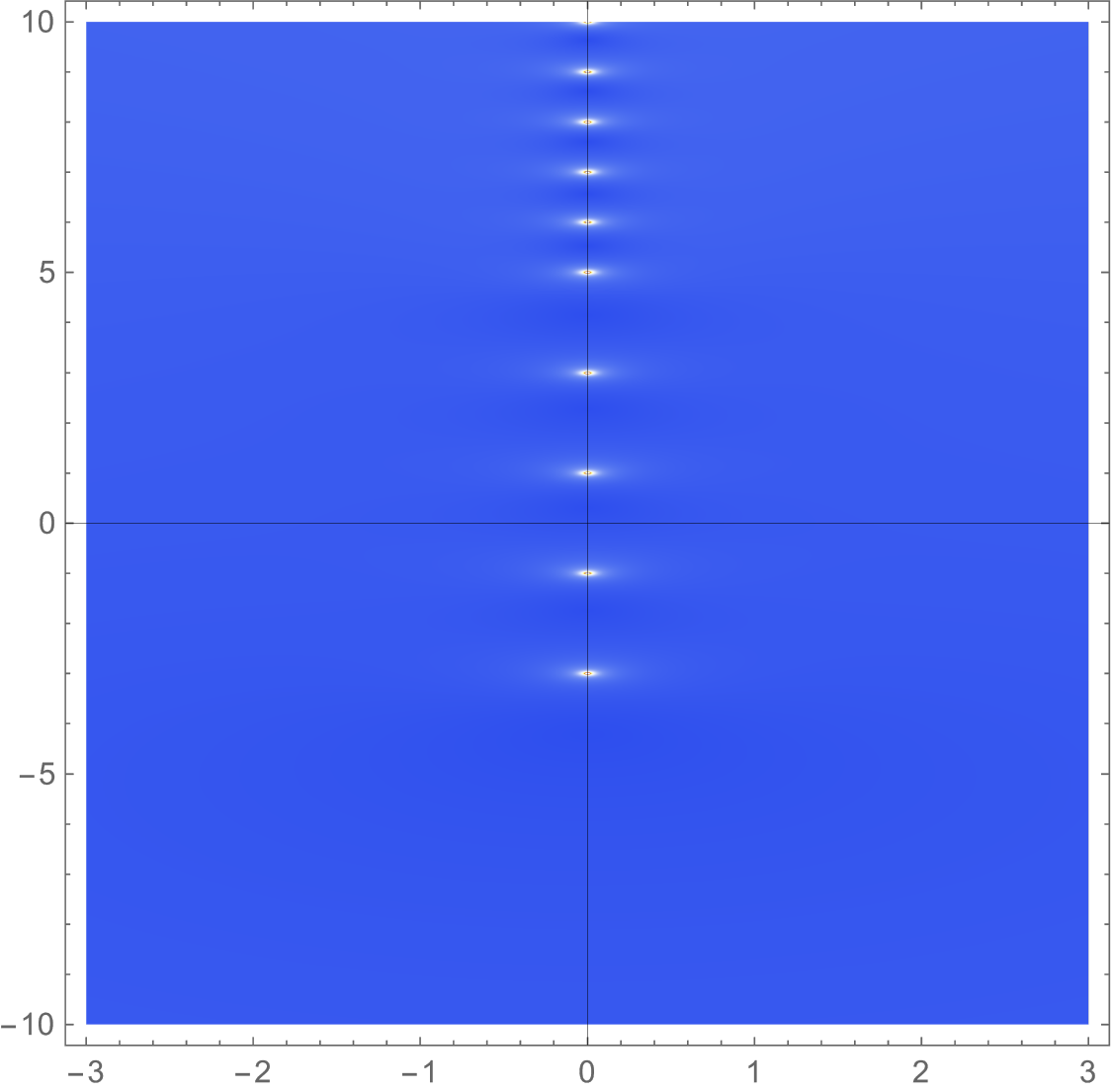}
            \caption[]%
            {{$\left|\Delta \rho^A_{l=0}(\omega)\right|$}}    
        \end{subfigure}
        \hspace{0.25cm}
        \begin{subfigure}{0.3\textwidth}  
            \centering 
            \includegraphics[height=4.0cm]{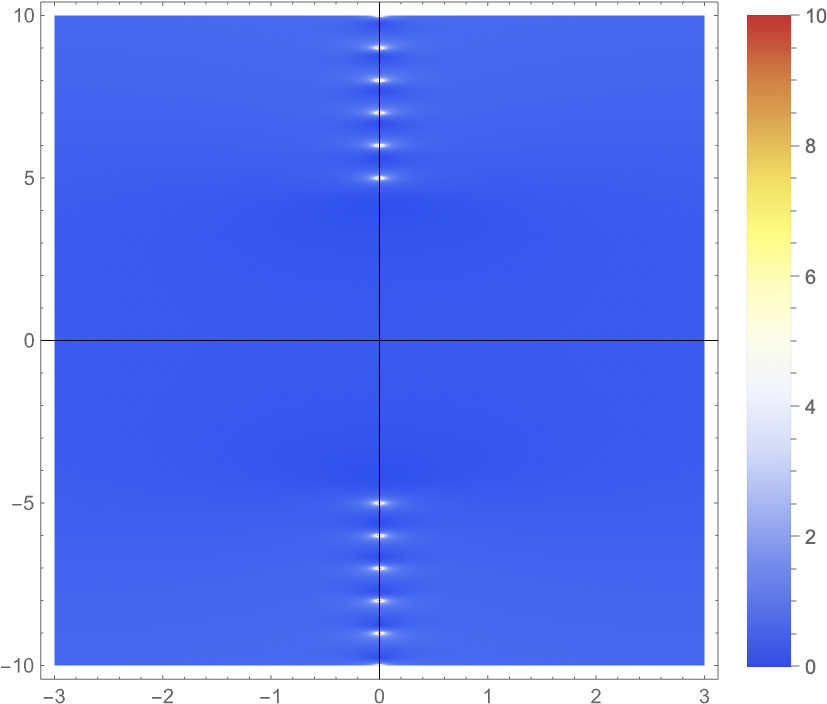}
            \caption[]%
            {{\small \centering  $\left|\Delta \rho_{l=0}(\omega)\right|$}}    
        \end{subfigure}
        \caption{Spectral densities for a exceptional scalar with $\Delta=-3$ in $dS_4$ in complex $\omega$-plane. { The white spots mark the poles.}}
        \label{pic:rhok3}
\end{figure}

In addition to a (anti-)resonance pole at $\omega=0$, in the shift-symmetric case there are a finite number of resonance poles lying on the {\it upper}-half plane, namely those with \eqref{eq:shiftedpoles}. By time reversal symmetry, we have the corresponding anti-resonance poles lying in the lower half plane. See figure \ref{pic:rhok3}.  
While the pole at $\omega=0$ always gets canceled in the total relative spectral density \eqref{eq:dossplit}, the poles \eqref{eq:shiftedpoles} do not, and will contribute to the character upon the inverse Fourier transform \eqref{eq:chiinverse}. For $t<0$, we close the contour in the lower half plane. This contour picks up all the poles of $\Delta\rho^R(\omega)$ except $\omega=0$ or those satisfying \eqref{eq:shiftedpoles}; in addition, it picks up the poles of $\Delta\rho^A(\omega)$ satisfying \eqref{eq:shiftedpoles}. Explicitly, we have
\begin{align}
    \int_{C_-} d
    \omega \, e^{-i\omega t}\tilde\rho^R(\omega) =&\frac{e^{kt}+e^{-(d+k) t}}{(1-e^{-t})^d}-1-\sum_{2n+l<k}D_l^d  e^{(k-2n-l)t} \;,\nn\\
    \int_{C_-} d
    \omega \, e^{-i\omega t} \tilde\rho^A(\omega) =&\sum_{2n+l<k}D_l^d  e^{-(k-2n-l)t} \; .
\end{align}
The $t<0$ case works similarly. Summing these contributions results in
\begin{align} \label{eq:tachyonCharacter}
    \chi(t) = \frac{e^{kt}+e^{-(d+k) t}}{|1-e^{-t}|^d}-1-\sum_{2n+l<k}D_l^d \left( e^{(k-2n-l)|t|} - e^{-(k-2n-l)|t|}\right) \;.
\end{align}
This gives a nice physical picture for the ``flipping" procedure in calculating the characters for the exceptional scalars \cite{Sun:2021thf}: there are a finite number of (anti-)resonance poles that lie in the upper (lower) half plane, and the subtraction by 1 accounts for the fact that the $\omega=0$ pole is never a pole of the total relative spectral density \eqref{eq:dossplit}.

To end this section, we note that the Harish-Chandra characters associated with (partially) massless fields \cite{Deser:1983tm,DESER1984396,Higuchi:1986py,Brink:2000ag,Deser:2001pe,Deser:2001us,Deser:2001wx,Deser:2001xr,Zinoviev:2001dt,Hinterbichler:2016fgl}, which include Maxwell field or graviton that are of great physical interests, exhibit a similar flipping structure \cite{Anninos:2020hfj,Sun:2021thf}. It would be interesting to check if the physical mechanism we just revealed for such a structure in the exceptional scalar case applies to the partially massless fields as well.

\section{Outlook}\label{sec:discussion}

We conclude by outlining several avenues for future exploration.

\paragraph{Spinning fields}

Our discussion has primarily revolved around free scalars. A natural extension involves exploring fields with spins, whose $S^{d+1}$ path integrals exhibit a distinct bulk-edge split \cite{Anninos:2020hfj} (extended to arbitrary static black holes in \cite{Grewal:2022hlo}). Indeed, elucidating the Lorentzian interpretation of this split originally motivated our work. The analysis is considerably more intricate and we will present the details in \cite{spin}.

\paragraph{More general backgrounds}

The reader might have noticed the robustness of the analysis in section \ref{sec:characterStaticPatchViewpoint}. In fact, it is easy to generalize it to free scalar theories on any (static) black holes, even in cases where the spectral functions have more complicated analytic structures such as branch cuts \cite{Ching:1995tj}. This is not too surprising, considering the parallel analyses for the partition functions that have been worked out in \cite{Law:2022zdq,Grewal:2022hlo}. 

\paragraph{Interacting theories and beyond}

Another natural next step is to generalize our considerations to interacting theories. In contrast to the free case, the spectral function in the interacting theory is no longer independent of the temperature, but it still plays a prominent role because all correlators can be written in terms of it by an integral representation \cite{Kapusta_Gale_2006,laine_basics_2016}. With the thermal correlators studied in this work, one can start performing perturbative calculations of various thermodynamic quantities. At the non-perturbative level, motivated by our key relation \eqref{eq:charGC}, one would naturally consider a character-like object
\begin{align}\label{eq:interchi}
    \chi(t) \equiv 2i \frac{d}{dt} \tilde \tr \, {\hat G}^C(t)\; , 
\end{align}
where 
\begin{align}\label{eq:interGC}
     G^C(t,x,\Omega|0,y,\Omega')\equiv  \Bigl\langle\left[ \hat{O}(t,x,\Omega),\hat{O} (0,y,\Omega')\right]\Bigr\rangle_{\beta=2\pi} 
\end{align}
is the spectral function for some scalar operator $\hat{O}$ in the {\it interacting} theory at temperature $\beta=2\pi$. Through a Källén–Lehmann-type representation \cite{Bros:1990cu,Loparco:2023rug} for \eqref{eq:interGC},\footnote{A Källén–Lehmann representation is discussed in the context of static patch in \cite{Mirbabayi:2022gnl}.} we can then write an analogous representation for \eqref{eq:interchi}
\begin{align}
    \chi(t)  = \int d\Delta \rho (\Delta) \chi_{\Delta}(t) \;. 
\end{align}
It would be interesting to explore the usefulness of organizing the physical information in this way.

Our novel perspectives on the correlators might also be inspiring for the bottom-up approach to a microscopic model associated with a static patch. Conceptually, consider the role of an {\it integrated} correlator as a notable group-theoretic object. This could offer a ``holistic" view that links together various proposals that posit microscopic degrees of freedom residing either on the stretched horizon \cite{Susskind:1993if,Shaghoulian:2021cef} or along the observer's worldline \cite{Anninos:2011af,Chandrasekaran:2022cip,Witten:2023xze}. 

If the microscopic Hilbert space is indeed finite-dimensional, its algebra of observables would be type-I \cite{Witten:2018zxz} for which notions such as traces and entropy are all well-defined. While one might not expect all these notions would survive in the semi-classical limit because of the change in the algebra type \cite{Witten:2021jzq}, it is possible that data such as relative traces or relative entropy remains intact. This makes objects like \eqref{eq:interchi} a natural candidate observable in a microscopic-macroscopic dictionary, encoding information of concepts like QNMs that emerge in the semi-classical limit \cite{Narovlansky:2023lfz,Parmentier:2023axg}. Another relevant comment is that the fact that \eqref{eq:interchi} is invariant under $T:t\to-t$ resonates with the idea that $T$ should be gauged in quantum gravity \cite{Harlow:2023hjb,Susskind:2023rxm}. We leave these speculative ideas for future investigation.

\section*{Acknowledgments} 

It is a great pleasure to thank Frederik Denef, Raghu Mahajan, Adel Rahman, Eva Silverstein and Douglas Stanford for stimulating conversations, and especially Dionysios Anninos, Varun Lachob, Klaas Parmentier, Zimo Sun and Andr\'{a}s Vasy for useful discussions and comments on the draft. AL  was supported in part by the Croucher Foundation, the Black Hole Initiative at Harvard University, the Stanford Science Fellowship, a Simons Investigator award, and NSF Grant PHY-2310429. MG was supported in part by the U.S. Department of Energy grant de-sc0011941.


\appendix

\section{Canonical quantization for free fields on the static patch}\label{sec:canquan}

We review the canonical quantization procedure for free scalars in a $dS_{d+1}$ static patch, which has been discussed in $d+1=4$ \cite{Higuchi:1986ww}, $d+1=2$ \cite{Akhmedov:2020qxd}, and later extended to general $d$ \cite{Akhmedov:2021cwh}.

Following section \ref{sec:staticLor}, a complete set of solutions to the Klein-Gordon equation is given by 
\begin{align}\label{appeq:fieldop}
    \phi(t,x,\Omega) = \sum_{l=0}^\infty \int_{-\infty}^\infty d\omega \, \phi_{\omega l}(t,x,\Omega) a_{\omega l} = \sum_{l=0}^\infty \int_{-\infty}^\infty d\omega \, e^{-i\omega t} \, C^{\rm n.}_{\omega l} \psi^{\rm n.}_{\omega l}(x) Y_l(\Omega) a_{\omega l} 
    \;,
\end{align}
where $\psi^{\rm n.}_{\omega l}(x)$ is the normalizable solution \eqref{eq:scalarSol}. The coefficients $C^{\rm n.}_{\omega l}$ are fixed by requiring 
\begin{align}\label{eq:KGnorm}
    \left( \phi_{\omega l} ,\phi_{\omega' l'}\right)_{\rm KG} = \delta(\omega-\omega')\delta_{l l'} \; , \qquad \omega, \omega'>0 \; , 
\end{align}
and 
\begin{align}\label{eq:KGnormneg}
     \left( \phi_{\omega l} ,\phi_{\omega' l'}\right)_{\rm KG} =- \delta(\omega-\omega')\delta_{l l'} \; , \qquad \omega, \omega'<0 \; ,
\end{align}
where 
\begin{align}
    \left( \phi_1 ,\phi_2\right)_{\rm KG} \equiv -i \int d^{d}x\sqrt{-g} g^{tt} \left(\phi_1\partial_t \phi^*_2 -\partial_t \phi_1 \phi^*_2 \right) 
\end{align}
is the Klein-Gordon inner product. Imposing these leads to the condition
\begin{align}\label{eq:KGcoecondition}
    2|\omega| |C^{\rm n.}_{\omega l}|^2 \int_0^\infty dx \,\bar\psi^{\rm n.}_{\omega l }(x)\bar\psi^{\rm n.}_{\omega' l }(x) =\delta(\omega-\omega')+\delta(\omega+\omega') \; , \qquad \omega ,\omega'\in \mathbb{R} \; , 
\end{align}
where $\bar\psi^{\rm n.}_{\omega l }(x)$ is the rescaled version of \eqref{eq:scalarSol}. To obtain this, we note that when $\omega$ and $\omega'$ have the same sign, \eqref{eq:KGnorm} and \eqref{eq:KGnormneg} implies \eqref{eq:KGcoecondition} without the second term on the right. To extend to the case when $\omega$ and $\omega'$ have opposite signs, observe that the left hand side of \eqref{eq:KGcoecondition} is invariant under $\omega \to -\omega$ or $\omega' \to -\omega'$; the term $\delta(\omega+\omega')$ is required so that both sides have this property.

From \eqref{appeq:fieldop}, we can also write down the mode expansion for its conjugate momentum,
\begin{align}
     \pi\equiv \frac{\partial \mathcal{L}}{\partial (\partial_t \phi)} = \sqrt{-g}g^{tt}\partial_t \phi \; . 
\end{align}
Upon quantization, \eqref{appeq:fieldop} is promoted to a self-adjoint field operator $\hat\phi$, with operator-valued coefficients $\hat a_{\omega l}$. Since $\psi^{\rm n.}_{\omega l}(x)$ is real, the self-adjointness of $\hat\phi$ implies 
\begin{align}
    C_{\omega l} = C^*_{-\omega l} \qquad \text{and} \qquad \hat a_{-\omega l}=\hat a^\dagger_{\omega l} \;.
\end{align}
The equal-time commutation relations are 
\begin{gather}
    \left[\hat\phi (t,x,\Omega),\hat \pi(t,y,\Omega') \right]= i\delta(x-y)\delta^{(d-1)}(\Omega,\Omega')\;, \nn\\
    \left[\hat\phi(t,x,\Omega),\hat \phi(t,y,\Omega') \right]=\left[\hat\pi(t,x,\Omega),\hat \pi(t,y,\Omega') \right]=0 \; ,
\end{gather}
which translate into the commutation relation 
\begin{gather}
   \left[\hat a_{\omega l}, \hat a_{\omega' l'}\right] =\sgn (\omega)\delta(\omega+\omega')\delta_{l l'}\; .
\end{gather}
The vacuum state is defined as one annihilated by $\hat a_{\omega l}$ for any $\omega>0$:
\begin{align}
     \hat a_{\omega l} \ket{0} = 0 \; , \qquad \forall \omega>0 \;, \quad l\geq 0 \; . 
\end{align}
By construction, such a (Boulware) vacuum is $SO(1,1)\times SO(d)$-invariant. This static vacuum is distinct from the global Euclidean vacuum which is $SO(1,d+1)$-invariant. For example, the expectation value of the stress tensor for a conformally coupled scalar in the static vacuum diverges at the horizon \cite{birrell_davies_1982}. Acting with creation operators $\hat a^\dagger_{\omega l}$ ($\omega>0$) on $\ket{0}$ then yields states labeled by occupation numbers $n_{\omega l}$ for each $(\omega, l)$ mode. On these, the actions of $\hat a_{\omega l}$ and $\hat a^\dagger_{\omega l}$ are
\begin{align}\label{eq:creationaction}
    \hat a_{\omega l} \ket{\cdots n_{\omega l} \cdots} = \sqrt{n_{\omega l}} \ket{\cdots n_{\omega l}-1 \cdots} \; , \quad \hat a^\dagger_{\omega l} \ket{\cdots n_{\omega l} \cdots} = \sqrt{n_{\omega l}+1} \ket{\cdots n_{\omega l}+1 \cdots} \; .
\end{align}
We end this appendix by recalling that the {\it quantum} retarded correlator in the static vacuum
\begin{align}
    i \bra{0} \left[\hat \phi (t,x,\Omega),\hat \phi (0,y,\Omega')\right] \ket{0} \theta(t)
\end{align}
satisfies the defining equations \eqref{eq:Greeneom} and \eqref{eq:GRcondition} for the retarded Green function associated with the Klein-Gordon equation, which is the reason why the analysis in section \ref{sec:causalfn} could be performed without reference to the quantum theory.

\section{Different correlators in thermal field theory}\label{sec:ThermalFT}

In this appendix we review some basics for the real-time and imaginary-time formalisms in thermal field theory \cite{Kapusta_Gale_2006,laine_basics_2016}. When we put the theory into a canonical ensemble, we are interested in computing the thermal averages of operators. We will use the notation
\begin{align}
    \langle \cdots \rangle_\beta \equiv \frac{1}{Z(\beta)}\Tr \left(e^{-\beta \hat H} \cdots \right) \; , \qquad Z(\beta) \equiv \Tr \, e^{-\beta \hat H} \;,
\end{align}
to denote the thermal average. In the following, we focus on bosonic operators $\hat O(t)$, which can be elementary fields or composite operators. For notational simplicity, we suppress other spatial dependence and quantum numbers such as spins.

\paragraph{Real-time correlators}

It is convenient to define the Wightman functions
\begin{align}\label{appeq:Wightman}
    G^>_{\beta}(t-t')\equiv\bigl\langle \hat O (t)\hat O (t')\bigr\rangle_\beta\; , \qquad   G^<_{\beta}(t-t')\equiv\bigl\langle \hat O (t')\hat O (t)\bigr\rangle_\beta \;, 
\end{align}
where we have assumed time translation invariance so that the functions only depend on $t-t'$. Without loss of generality we will set $t'=0$ from now on. Using 
\begin{align}
    e^{-\beta \hat H}\hat O (t) e^{\beta \hat H} = \hat O (t+i\beta ) 
\end{align}
and the cyclicity of the trace, one can derive the Kubo-Martin-Schwinger (KMS) relation
\begin{align}
    G^>_{\beta}(t)=G^<_{\beta}(t+i\beta) \; , 
\end{align}
or in frequency space,
\begin{align}\label{eq:KMS}
    \mathcal{G}^>_\beta (\omega) = e^{\beta \omega} \mathcal{G}^<_\beta (\omega) \; .
\end{align}
Starting from \eqref{appeq:Wightman}, one can construct all other correlators by taking (anti-)commutators and multiplying the step functions $\theta(\pm t)$. A simple one is to take the average of \eqref{appeq:Wightman} to obtain the symmetric Wightman function or statistical correlator:
\begin{align}
    G^{\rm Sym}_{\beta}(t)\equiv\frac12\bigl\langle \hat O (t)\hat O (0)+\hat O (0)\hat O (t)\bigr\rangle_\beta \; .
\end{align}
It is sometimes useful to study the so-called 2-sided correlator:
\begin{align}\label{eq:2sided}
    G^{12}_\beta (t) = G^>_{\beta}\left(t-\frac{i \beta}{2}\right) \; ,
\end{align}
{which can be understood as the cross-copy correlator in the
thermofield-double (TFD) purification of the thermal state,
\begin{equation}
    G^{12}_\beta(t)
    = \bra{\mathrm{TFD}} \hat O_L(t)\,\hat O_R(0) \ket{\mathrm{TFD}}\; ,
    \qquad
    \ket{\mathrm{TFD}}
    \equiv \frac{1}{\sqrt{Z(\beta)}} \sum_E e^{-\beta E/2}\, \ket{E}_L \otimes \ket{E}_R \;,
\end{equation}
where $\hat O_L \equiv \hat O \otimes I $ and $\hat O_R \equiv I \otimes \hat O^T $.
}

In linear response theory, the retarded and advanced propagators,
\begin{align}
    G^R_{\beta}(t)\equiv i \Bigl\langle \left[\hat O (t),\hat O (0)\right]\Bigr\rangle_\beta \theta(t)\; , \qquad   G^A_{\beta}(t)\equiv - i \Bigl\langle \left[\hat O (t),\hat O (0)\right]\Bigr\rangle_\beta \theta(-t) \;,
\end{align}
often arise. Closely related is the so-called spectral or commutator function,
\begin{align}
    G^C_{\beta}(t) \equiv \Bigl\langle \left[\hat O (t),\hat O (0)\right]\Bigr\rangle_\beta \;,
\end{align}
which plays a distinguished role in section \ref{sec:characterStaticPatchViewpoint}. 

In frequency space, the Wightman functions are related to the spectral function by{
\begin{align}\label{eq:wightman}
    \mathcal{G}^>_\beta (\omega) =  \frac{1}{1-e^{-\beta \omega}}\mathcal{G}^C_\beta (\omega) \; , \qquad 
    \mathcal{G}^<_\beta (\omega) = \frac{1}{e^{\beta \omega}-1} \mathcal{G}^C_\beta (\omega) \; , 
\end{align}}
implying 
\begin{align}\label{eq:GsymGcrelation}
    \mathcal{G}^{\rm Sym}_{\beta}(\omega)= { \frac12} \coth \frac{\beta \omega}{2}\mathcal{G}^C_{\beta}(\omega) \; .
\end{align}
We also have
\begin{align}
    \mathcal{G}^C_{\beta}(\omega) =-i\left(\mathcal{G}^R_{\beta}(\omega) -\mathcal{G}^A_{\beta}(\omega) \right)\;, 
\end{align}
and
\begin{align}
    \mathcal{G}^{12}_l (\omega) = \frac{\mathcal{G}^C_l (\omega)}{ 2\sinh  \frac{\beta \omega}{2}}\; . 
\end{align}

\paragraph{Euclidean correlators}

In non-perturbative formulations, it is convenient to define a correlator in the imaginary time $0\leq \tau \leq \beta$:
\begin{align}
        G^E_{\beta}(\tau)=\bigl\langle \hat O (-i\tau)\hat O (0)\bigr\rangle_\beta \; .
\end{align}
The KMS condition is equivalent to the periodicity of the Euclidean correlator:
\begin{align}
    G^E_{\beta}(\tau)=G^E_{\beta}(\tau+\beta) \; . 
\end{align}
We can expand this into a Fourier series,
\begin{align}
    G^E_{\beta}(\tau) = \frac{1}{\beta}\sum_{k\in \mathbb{Z}}e^{-i \omega^{\rm Mat}_{k} \tau}\mathcal{G}^E_{\beta}(\omega_k) \;,
\end{align}
where $\omega^{\rm Mat}_{k}$ are the so-called Matsubara frequencies
\begin{align}\label{appeq:matfreq}
    \omega^{\rm Mat}_{k} = \frac{2\pi k}{\beta} \; , \qquad k \in \mathbb{Z} \; .
\end{align}

\paragraph{Free scalars on the static patch}

For the discussion of this paper, we work out some formulas for the case of a free scalar in the static patch. Plugging (the quantized version of) \eqref{appeq:fieldop} into the Wightman function \eqref{appeq:Wightman} and using \eqref{eq:creationaction}, one can show 
\begin{align}\label{eq:Wightmanfree}
    G^>_{\beta}(t;x,\Omega|y,\Omega') 
     = \sum_{l=0}^\infty \int_0^\infty d\omega \left[n_B (\omega)  \, e^{i \omega t}+\left(n_B (\omega)+1\right) \, e^{-i \omega t}  \right] |C^{\rm n.}_{\omega l}|^2 \psi^{\rm n.}_{\omega l} (x)\psi^{\rm n.}_{\omega l} (y)Y_l(\Omega)Y_l(\Omega')\; . 
\end{align}
We have set $t_x = t, t_y =0$. The factor
\begin{align}\label{appeq:Bose-Einstein}
    n_B ( \omega) \equiv \frac{1}{e^{\beta \omega}-1}
\end{align}
is the Bose-Einstein distribution. Similarly, $G^<_{\beta}(t;x,\Omega|y,\Omega')$ is given by \eqref{eq:Wightmanfree} but with $t\to-t$. Taking the difference, it is easy to see that the Bose-Einstein factor \eqref{appeq:Bose-Einstein} drops out, implying that the spectral function, and thus the retarded and advanced functions, are independent of the temperature, and therefore exactly equal to that at zero temperature:
\begin{align}\label{appeq:Gc}
    G^C_{\beta}(t;x,\Omega|y,\Omega') = &G^C_{\beta=\infty}(t;x,\Omega|y,\Omega')\nn\\
    =& \sum_{l=0}^\infty \int_{0}^\infty d\omega \,  \left(e^{-i\omega t}-e^{i\omega t}\right)|C^{\rm n.}_{\omega l}|^2 \psi^{\rm n.}_{\omega l} (x)\psi^{\rm n.}_{\omega l} (y)Y_l(\Omega)Y_l(\Omega')\; . 
\end{align}
Taking the average of \eqref{eq:Wightmanfree} and $G^<_{\beta}(t;x,\Omega|y,\Omega')$, one can see that \eqref{eq:GsymGcrelation} is satisfied.

Before we end, we note that by comparing \eqref{eq:spectralfnformula} with the mode expansion \eqref{appeq:Gc}, we can read off the normalization such that the condition \eqref{eq:KGcoecondition} is satisfied: 
 \begin{align}
     \left| C^{\rm n.}_{\omega l} \right|^2 = C^{\rm n.}_{\omega l}C^{\rm n.}_{-\omega l}= \frac{|\omega|}{\pi \, W^{\rm Jost}_l(\omega)\,W^{\rm Jost}_l(-\omega)} \;.
 \end{align}


\section{Spectral theory in 2D Rindler space}\label{app:Rin2D}

In this section, we study the scattering problem 
\begin{align}\label{appeq:rindlerscatt}
	\left[ -\partial_{x}^2 + \left(\frac{4\pi}{\beta} \right)^2 e^{-\frac{4\pi}{\beta}(x-a)} \right] \psi(x) = \omega^2 \psi(x) \; ,
\end{align}
associated with the Klein-Gordon equation (with mass $m^2=\left(\frac{4\pi}{\beta} \right)^2 e^{\frac{4\pi a}{\beta}}$) on the 2D Rindler space, 
\begin{align}\label{appeq:refM}
     ds^2 = e^{-\frac{4\pi}{\beta}x} \left( -dt^2+dx^2 \right) \; , \qquad -\infty<x<\infty \; ,
\end{align}
with $x\to -\infty$ and $x\to \infty$ corresponding to the asymptotic infinity and horizon respectively. The Klein-Gordon equation on any static non-extremal black hole reduces to \eqref{appeq:rindlerscatt} near the horizon, with $\beta$ set by the black hole temperature and $a$ depending on the transverse quantum numbers and the mass. For example, \eqref{eq:nearhordsKG} corresponds to $\beta=2\pi$ and $a=\frac12\log \left( \Delta+l-1\right)\left( \bar\Delta+l-1\right)$.

The problem has been partially analyzed before in \cite{Law:2022zdq}. The normalizable solution to \eqref{appeq:rindlerscatt} is  
\begin{align}
 \psi_\omega^\text{n.}  (x) = K_{\frac{i\beta \omega}{2\pi} }\left( 2\,e^{-\frac{2\pi }{\beta} (x-a)}\right) \; ,
\end{align}
which exponentially decays as $x \to -\infty$:
\begin{align}
	K_{\frac{i\beta \omega}{2\pi} }\left( e^{-\frac{2\pi }{\beta} (x-a)}\right)  \propto e^{\frac{\pi }{\beta} (x-a)- 2\,e^{-\frac{2\pi }{\beta}(x-a)}}\; .
\end{align}
Near the Rindler horizon, it is a linear combination of plane waves:
\begin{align}\label{eq:Rindler near hor}
		\psi_\omega^\text{n.}  (x)\stackrel{x\to \infty}{\approx}\frac12 \, \Gamma
	\left(\frac{i\beta \omega}{2\pi}\right) e^{i \omega  (x-a)}+\frac12\,\Gamma
	\left(-\frac{i\beta \omega}{2\pi} \right) e^{-i \omega  (x-a)} \; .
\end{align}
We also obtain the Jost solution 
\begin{align}\label{eq:jostsolRin}
	\psi_\omega^{\rm Jost}(x)=   e^{i \omega a} \Gamma \left(1-\frac{i \beta  \omega }{2 \pi }\right) I_{-\frac{i \beta  \omega }{2 \pi }}\left(2\, e^{-\frac{2 \pi  }{\beta
	}(x-a)}\right)\;,
\end{align}
which is purely ingoing at the horizon:
\begin{align}
	\psi_\omega^{\rm Jost}(x)\stackrel{x\to \infty}{\approx} e^{i \omega x} \; .
\end{align}
The ratio between the coefficients of the outgoing and incoming waves of the normalizable mode in \eqref{eq:Rindler near hor} defines a unitary S-matrix:
\begin{align}\label{appeq:RindlerS}
	\mathcal{S}^\text{Rin} (\beta,a,\omega) = e^{-2i \omega a}\frac{\Gamma
		\left(\frac{i \beta  \omega }{2 \pi } \right)}{\Gamma
		\left(-\frac{i \beta  \omega }{2 \pi } \right)} \; .
\end{align}
The retarded Green function for \eqref{appeq:rindlerscatt} is given by \cite{skinner2014mathematical}
\begin{align}
	2\pi \mathcal{G}^R(\omega ; x|y ) =\frac{\psi_\omega^\text{n.}  (x)\psi_\omega^{\rm Jost}(y) }{W^{\rm Jost}(\omega)}\theta(y-x)+\frac{\psi_\omega^\text{n.}  (y)\psi_\omega^{\rm Jost}(x) }{W^{\rm Jost}(\omega)}\theta(x-y)\; , 
\end{align}
where the Jost function can be computed to be 
\begin{align}\label{eq:WronRind}
 	W^{\rm Jost}(\omega)
 = e^{i \omega a} \frac{2\pi}{\beta}\Gamma \left(1-\frac{i \beta  \omega }{2 \pi }\right)  \; .
\end{align}
The spectral function is given by the difference
\begin{align}
    \mathcal{G}^C(\omega ; x|y ) =-i\left(\mathcal{G}^R(\omega ; x|y ) -\mathcal{G}^R(-\omega ; x|y ) \right)\; .
\end{align}
Using the fact that $K_{i\lambda} (z)=K_{-i\lambda} (z), \lambda \in \mathbb{R}$ and the identity
\begin{align}
	I_{i\lambda} (z)-I_{-i\lambda} (z)=-\frac{2 i \sinh (\pi  \lambda )}{\pi } K_{i\lambda} (z) \; ,
\end{align}
we compute 
\begin{align}
	\mathcal{G}^C(\omega ; x|y )=\frac{\beta}{2\pi^3}\sinh \frac{\beta\omega}{2} \psi_\omega^\text{n.}  (x) \psi_\omega^\text{n.}  (y) \;.
\end{align}


\bibliographystyle{utphys}
\bibliography{ref}

\end{document}